\documentclass[showpacs,aps,prb,twocolumn,superscriptaddress]{revtex4-2}

\usepackage[english]{babel}
\usepackage[T1]{fontenc}
\usepackage[utf8]{inputenc}
\usepackage{amsmath}
\usepackage{amsthm}
\usepackage{amsfonts}
\usepackage{xcolor}
\usepackage{xr-hyper}
\usepackage{hyperref}
\usepackage{amssymb,graphicx}
\usepackage{tikz}
\usetikzlibrary{decorations.pathmorphing}
\usepackage{circuitikz}
\usepackage{blochsphere}
\usepackage{siunitx}
\usepackage{mathbbol}
\usepackage{amsmath}    
\usepackage{braket}
\usepackage{verbatim}
\usepackage{tabularx}

\newcommand{\Figref}[1]{Fig.~\ref{#1}}
\newcommand{\Eqref}[1]{Eq.~(\ref{#1})}
\newcommand{\Apref}[1]{appendix~\ref{#1}}

\newsavebox\MBox

\newcommand*{\colorboxed}{}
\def\colorboxed#1#{%
  \colorboxedAux{#1}%
}
\newcommand*{\colorboxedAux}[3]{%
  \begingroup
    \colorlet{cb@saved}{.}%
    \color#1{#2}%
    \boxed{%
      \color{cb@saved}%
      #3%
    }%
  \endgroup
}

\begin{document}
%
%
\title{Optimal Connectivity from Idle Qubit residual coupling Cross-Talks\texorpdfstring{\\in a Cavity Mediated Entangling Gate}{}}
%
%
\author{A.~Mammola}
\affiliation{C12 Quantum Electronics, Paris, France}
\author{M.~M.~Desjardins}
\affiliation{C12 Quantum Electronics, Paris, France}
\author{Q.~Schaeverbeke }
\affiliation{C12 Quantum Electronics, Paris, France}
\date{\today}
%
\begin{abstract}
Quantum processors operated through long range interaction mediated by a microwave resonator have been envisioned to allow for high connectivity.
The ability to selectively operate qubits rely on the possibility to dynamically suppress the coupling between each qubit and the resonator, however there always remains a residual coupling.
In this article, we investigate the effect of high processor connectivity on average two qubit gate fidelity in a cavity based architecture with tunable coupling.
Via a perturbative approach, we quantify the cross-talk errors from transverse residual couplings and show that they scale as $nm^2$ where $n$ is the number of idle qubits and $m$ is the ratio between the transverse residual and active couplings.
Setting an error threshold $E_\mathrm{thr}$, we demonstrate that cross-talks restrict the hardware topology and prevent the full use of all-to-all connectivity.
We predict that the maximum number of qubits allowed by $E_\mathrm{thr}$ scales as $n \propto E_\mathrm{thr}/m^2$.

\end{abstract}

\maketitle

\section{Introduction}
Obtaining the strong coupling regime between a single atom and a photon has been a key achievement in studying light matter interaction at the atomic scale.
The advent of cavity quantum electrodynamics (QED) \cite{haroche1989}, by concentrating an electric field in a small region of space around an atom, has allowed to reach this regime and has become the test-bed to study models such as the Rabi and Tavis-Cummings models \cite{tavis1968}.
Naturally this light matter interaction has been considered as a way to produce and control entanglement between two atoms \cite{chilingaryan2013}.
The principles of cavity QED have then been adapted to quantum circuits made of semiconducting or superconducting components in a field now known as circuit QED \cite{cottet2015,blais2021}, in which an artificial atom is made out of circuit components, allowing for better control of its features.
This cavity mediated qubit interaction has the advantage of being non-local and to be able to propagate entanglement on a large number of qubits as shown in the Tavis-Cummings model \cite{song2017}.
Such long range interaction could be used in a quantum computer to reach very high hardware connectivity leading to more efficient entanglement propagation \cite{hazra2021}.
To evaluate how feasible all-to-all connectivity is in practice, we have to explore the scaling of cavity based architectures.
For quantum processors, scaling remains a major challenge in any technology and cross-talks have been identified as a key obstacle {\cite{monroe2013,xia2015,saffman2016,krinner2020,heinz2021,kandala2021,zhao2022,wei2022,undseth2023,zhou2023,heinz2024}.
Cross-talks arise from two phenomena.
Leakage of single qubit driving field to neighboring qubit or unwanted interactions between qubits \cite{Sarovar2020} (''sometimes referred as quantum cross-talks``), resulting in the deterioration of the gate fidelities due to coherent errors.
In this article we focus on the latter and the term ''cross-talk`` will always refer to residual couplings between qubits.
The use of a long-range interaction through a microwave cavity allows theoretically for all-to-all connectivity, if one is able to suppress these unwanted qubit interactions.
Indeed, in the dispersive coupling regime between the qubits and the cavity, the entanglement of two qubits is mediated by a virtual photon exchange via a microwave resonator \cite{benito2019}.
Hence, as soon as two qubits are coupled to the resonator, they can be entangled independently of their spatial distribution in the hardware.
Connectivity can be critical for a quantum hardware as it drastically impacts the algorithm's compilation.
Low connectivity induces a significant overhead in algorithms that require the propagation of entanglement along a qubit register \cite{matsuo2019,Holmes_2020,hashim2022}.
Two distant qubits that need to be entangled will have to be swapped until made close to each other, increasing the circuit depth by the amount of required swaps \cite{Qubit_routing_prob}.
High connectivity, on the other hand, requires few intermediate swaps, but implies being able to decouple any qubit from the resonator.
For example, in the case of cavity coupled semiconducting qubits, this can be done through electrostatic control of the qubits, however the suppression of the coupling is never perfect.
The residual coupling between the idle qubits and the resonator produces cross-talk errors that stack up with the total number of qubits.
The suppression of unwanted couplings can be assisted using dynamical decoupling and some pulse sequences have already been proposed for the suppression of longitudinal residual couplings for superconducting single qubit gates \cite{tripathi2022, zhou2023, ezzell2023, niu2024} and singlet-triplet semiconducting spin qubit single qubit gates \cite{buterakos2018}.
In this article we quantify the error coming from the transverse cross-talk errors on an $i\rm{SWAP}$ gate, native in cavity based processors \cite{schuch2003,ni2018,benito2019,warren2019,sung2021,dijkema2023}, for hardwares with tunable cavity-qubit coupling.
Predicting the scaling of the cross-talk errors with the number of qubits connected through the same resonator, we deduce an optimal number of qubits for a single resonator based on current quantum error-correction threshold.
We believe that properly quantifying the effect of such cross-talk errors is paramount in guiding the design of the hardware towards its optimal connectivity.
Of course other constraints, such as the effect on the resonator quality factor of stacking circuit components near it for example, have to be explored to make a final decision on the hardware design.
We do not consider dynamical decoupling as, to our knowledge, no  scheme has been proposed to address transverse cross-talk errors and no analytical estimate of the cross-talk error has been proposed so far.
Hence, we focus in this article on the $XX$ errors provoked by idle qubits during two qubit gates in an all-to-all connected hardware.
Future work will have to investigate dynamical decoupling schemes capable of reducing transverse cross-talk errors.
The article is organized as follows.
We introduce the model of the cavity mediated entangling gate in \autoref{section:Ham}.
The dynamics of the $i\mathrm{SWAP}$ gate is presented and we deduce the effect of idle qubits.
In \autoref{section:models} we present the framework used to analyse the cross-talk errors.
We predict the scaling of the error rate with the number of idle qubits which leads us to predicting the maximum number of qubits that can be used in a single resonator with respect to an error threshold that we set.
Such a prediction can strongly affect the topology of a processor.
Numerical comparisons are presented as a reference using QuTiP simulations and will be used as a benchmark for our analytical calculations. 
Finally, \autoref{section:concl} provides a conclusion on this work with perspectives.
Throughout this article, we will use $\hbar=1$.
\section{Cavity mediated entanglement}\label{section:Ham}
\subsection{Ideal two qubit gate}
We assume, for simplicity, that we are able to produce reproducible qubits that all interact with the same resonator.
While in practice the assumption of reproducible qubits is challenging, relaxing it would only produce minor effects that do not impact the argument we make through this article.
We also assume that we are able to dynamically tune their coupling intensities with the resonator independently \cite{Cottet2010, gambetta2011}.
Using this property we are able to target which qubits are involved in a two qubit gate.
Given $N$ qubits, in a frame rotating around each of the qubit's quantization axis at a speed defined by their frequencies, we write the cavity--qubit Hamiltonian as \cite{dicke1954,cottet2015}
\begin{equation}\label{eq:H}
    H=\sum_{i=1}^{N}\frac{\omega_i}{2}\sigma_z^{(i)}+\omega_ca^\dag a+\sum_{i=1}^{N}g_i\sigma_x^{(i)}(a+a^\dag)
\end{equation}
where $a^\dag$ is the creation operator of a mode of frequency $\omega_c$ in the cavity, $\sigma_j^{(i)}$ are the Pauli operator acting on the qubit $i$ and $g_i$ is the qubit $i$ cavity coupling.
For two dispersively coupled qubits with the cavity, meaning that $g_i\ll|\omega_c-\omega_i|$, we can show that the qubits are able to exchange an excitation through a virtual photon exchange with the cavity, as depicted in Fig. \ref{fig:spinq}.
Assuming the resonator remains in its ground state during the operation, one shows that \Eqref{eq:H} for two qubits is equivalent to \cite{benito2019,warren2019}
\begin{equation}\label{hswap}
    H_\mathrm{2q}=H_0+\gamma H_\mathrm{S},
\end{equation}
where
\begin{equation}
    \left\{
    \begin{aligned}
        &H_0=\sum_{i=1}^{N}\frac{\omega_i}{2}\sigma_z^{(i)}\\
        &H_\mathrm{S}=\sigma_+^{(1)}\sigma_-^{(2)}+\sigma_-^{(1)}\sigma_+^{(2)}.
    \end{aligned}
    \right.
\end{equation}
The dependence of $\gamma$ with the parameters defining the qubit is deduced from the transformaton of \Eqref{eq:H} into \Eqref{hswap} and reads \cite{Economu}
\begin{equation}\label{eq:gamma}
    \gamma=g_1g_2\omega_c\left(\frac{1}{\omega_1^2-\omega_c^2}+\frac{1}{\omega_2^2-\omega_c^2}\right),
\end{equation}
From \Eqref{hswap} we can deduce the unitary evolution of the qubit.
Assuming that both qubits are identical at the \textit{on} point, in a frame rotating at frequency $\omega=\omega_1=\omega_2$ around the $z$ axis of the qubits \cite{benito2019}
\begin{equation}
    U_\mathrm{S}=\begin{pmatrix}
        1&0&0&0\\
        0&0&i&0\\
        0&i&0&0\\
        0&0&0&1
    \end{pmatrix}
\end{equation}
where we have set the gate time $t_g=\pi/2\gamma$.
This unitary will be used later on as a reference for evaluating the cross-talk errors.
\begin{figure}
    \centering
    \resizebox{\columnwidth}{!}{%
    \begin{tikzpicture}

    \fill[red!10!white] (0,0)rectangle(3,2);
    \fill[blue!10!white] (0.75-0.1,1)circle(0.35);
    \fill[blue!10!white] (2.25+0.1,1)circle(0.35);

    \draw[fill=black!20] (0,0.3) .. controls (0.5,0.1) and (2.5,0.1) .. (3,0.3)--(3,0)--(0,0)--(0,0.3);

    \draw[fill=black!20] (0,0.7+1) .. controls (0.5,0.9+1) and (2.5,0.9+1) .. (3,0.7+1)--(3,1+1)--(0,1+1)--(0,0.7+1);

    \draw[blue, line width=0.75pt] (0.75-0.1,1)circle(0.35);
    \fill[red] (0.75-0.1,1.35)circle(0.075);
    \fill[blue] (0.75-0.1,0.65)circle(0.075);
    \draw[blue, line width=0.75pt] (2.25+0.1,1)circle(0.35);
    \fill[blue] (2.25+0.1,1.35)circle(0.075);
    \fill[red] (2.25+0.1,0.65)circle(0.075);

    \begin{scope}
        \clip (0,1)rectangle(3,0);
        \draw[blue, line width=0.75pt] (0.75-0.1,1)ellipse(0.35 and 0.125);
    \end{scope}
    
    \draw[blue, line width=0.75pt, densely dashed] (0.75-0.1,1)ellipse(0.35 and 0.125);

    \begin{scope}
        \clip (0,1)rectangle(3,0);
        \draw[blue, line width=0.75pt] (2.25+0.1,1)ellipse(0.35 and 0.125);
    \end{scope}
    
    \draw[blue, line width=0.75pt, densely dashed] (2.25+0.1,1)ellipse(0.35 and 0.125);

    \draw[latex-, red, line width=0.8pt] (0.65-0.1,0.55)arc(262:98:0.45);
    \draw[-latex, red, line width=0.8pt] (0.85+1.5+0.1,0.55)arc(-82:88:0.45);

    \draw[latex-latex, red, line width=0.8pt,decorate, decoration={snake, segment length=1mm, amplitude=1mm, pre length=2mm, post length=2mm}] (1.05,1)--(1.95,1);
    
\end{tikzpicture}
    }
    \caption{Schematic representation of the cavity mediated entangling gate. Two qubits, represented as two blue Bloch spheres, are coherently interacting with the electric field (in light red) of a resonator (in grey). As the qubits interact through the cavity a direct exchange of the population occurs giving rise to an $i\mathrm{SWAP}$ gate. The red dots represent the initial state of the qubits while the blue dots represent there final states.}\label{fig:spinq}
\end{figure}
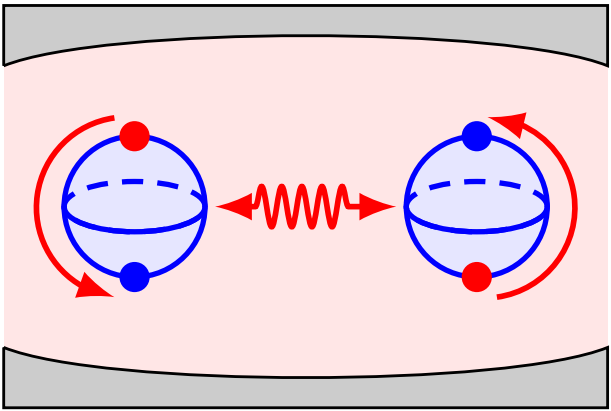
\subsection{Idle qubits}
From the qubit coupling mechanism, one understands that as soon as an additional qubit is connected to the the cavity, it will also be able to take part in the virtual photon exchange which produce a delocalised state on all connected qubits.
Let $N$ qubits be embedded in the same microwave resonator as depicted in \Figref{fig:idle}.
Two of these qubits are set active at the \textit{on} point and the other $n=N-2$ qubits are set idle at the \textit{off} point.
Practically, the way qubits are set \textit{on} or \textit{off} depends on the nature of the qubit.
For semiconducting spin qubit a definition that we will conform to in this article is given in \cite{Cottet2010}.
Single electron semiconducting qubits are encoded in DQD for which the chemical potential bias between each dots determines if the qubit is active or idle as the bias has a direct impact on the qubit frequency and its coupling to the cavity \cite{Cottet2010,burkard2021,dijkema2023} (this is the case depicted in \autoref{A1} in \Figref{sup:fig1}).
This effect can also be obtained tuning the tunnel barrier of the DQD instead \cite{benito2019}.
Hence, to put it more simply, the \textit{on} and \textit{off} points are two different points on the (chemical potential bias, tunnel rate) 2D manifold of each individual qubits.
This allows for independent and fast setting of all qubits in an active (\textit{on}) or a passive (\textit{off}) state.
Note however that for each individual qubit, its frequency and cavity coupling are jointly affected by this manipulation.
Superconducting qubits have a different way of being manipulated \cite{blais2021}, commonly they rely on controlling the flux in a SQUID \cite{kjaergaard2020,blais2021}.
Depending on the way the qubit is made, this gives control over the qubit frequency \cite{majer2007} or over its coupling to the cavity \cite{gambetta2011, bialczak2011, chen2014, neill2018}.
We focus in our study on the latter, where the coupling to the cavity is tunable.
In this case, some proposals have been made towards all-to-all connectivity \cite{hazra2021}.
Choosing the same \textit{off} and \textit{on} point for all qubits, we can expect that as long as two qubits are in the same regime, they will be equivalent and exhibit the same properties.
Usually in superconducting processors, one aims at choosing different \textit{off} points for idle qubits, in order to reduce the energy exchange among them, however, having tunable cavity-qubit couplings allows to have negligible idle qubit interactions and relaxes the requirement of having different frequencies for all qubits.
This should make it easier to scale the quantum processor as we are not as impacted by frequency crowding in this situation.

We define $\gamma$ as the coupling between two active qubits, $\gamma^\prime$ as the coupling between an idle qubit and an active one.
One can show (see \Apref{A1}) that the Hamiltonian for the ensemble of $N$ qubits dispersively coupled to a resonator is
\begin{equation}\label{Heff}
    H=\gamma H_\mathrm{S}+\gamma^\prime H^\prime+\tilde{\gamma} \tilde{H} + \tilde{\omega} H_{zz}
\end{equation}
where $\tilde{\omega}$ is a function of the nature (semiconducting or superconducting) and parameters of the qubit.
For semiconducting qubits it is non-zero only for idle qubits.
\begin{equation}
    H^\prime=\sum_{i=1}^2\sum_{j=3}^{N}\left(\sigma_{-}^{(i)}\sigma_{+}^{(j)} +\sigma_{+}^{(i)}\sigma_{-}^{(j)}\right)
\end{equation}
is the transverse interaction Hamiltonian between the \textit{on} qubits and the \textit{off} qubits,
\begin{equation}
        \tilde{H}=\sum_{i=3}^{N-1}\sum_{j=i+1}^{N}\left(\sigma_{-}^{(i)}\sigma_{+}^{(j)} +\sigma_{+}^{(i)}\sigma_{-}^{(j)}\right),
\end{equation}
is the transverse interaction Hamiltonian between the \textit{off} qubits,
and
\begin{equation}
    H_{zz}=\sum_{i=3}^{N-1}\sum_{j=i+1}^N\sigma_{z}^{(i)}\sigma_{z}^{(j)}
\end{equation}
is the longitudinal interaction Hamiltonian between the \textit{off} qubits.
This type of Hamiltonian is reminiscent of the quantum anisotropic Curie-Weiss model \cite{chayes2008,bulekov2021}, an Ising model on a complete graph that has a natural implementation for the type of hardware targeting all-to-all connectivity.
In the following we assume that the \textit{on} photon--qubit coupling is much larger than the \textit{off} photon-qubit coupling such that $\gamma\gg\gamma^\prime\gg\tilde{\gamma}\sim\tilde{\omega}$, as they respectively scale as $g_\textit{on}^2$, $g_\textit{on}g_\textit{off}$ and $g_\textit{off}^2$ according to \Eqref{eq:gamma} with $g_\textit{on}\gg g_\textit{off}$.
In that case the longitudinal coupling $\tilde{\omega}$ and the idle-idle coupling $\tilde{\gamma}$ are dominated by the transverse couplings $\gamma$ and $\gamma^\prime$ and we neglect the contributions of $\tilde{H}$ and $H_{zz}$, wimplifying our Hamiltonian into
%
%
\begin{equation}
    H=\gamma H_\mathrm{S}+\gamma^\prime H^\prime,
\end{equation}
in which the term $\gamma^\prime H^\prime$ is considered to be a perturbation of $H_S$.
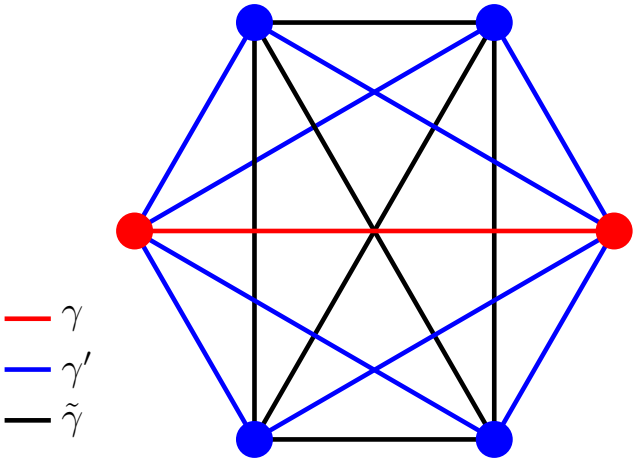
\begin{figure}
    \centering
    \resizebox{\columnwidth}{!}{%
    \begin{tikzpicture}
    \draw[line width = 0.5mm] (0.5*0.3\columnwidth,0.87*0.3\columnwidth)--(-0.5*0.3\columnwidth,0.87*0.3\columnwidth);
    \draw[line width = 0.5mm] (0.5*0.3\columnwidth,0.87*0.3\columnwidth)--(0.5*0.3\columnwidth,-0.87*0.3\columnwidth);
    \draw[line width = 0.5mm] (0.5*0.3\columnwidth,0.87*0.3\columnwidth)--(-0.5*0.3\columnwidth,-0.87*0.3\columnwidth);
    \draw[blue, line width = 0.5mm] (0.5*0.3\columnwidth,0.87*0.3\columnwidth)--(0.3\columnwidth,0);
    \draw[blue, line width = 0.5mm] (0.5*0.3\columnwidth,0.87*0.3\columnwidth)--(-0.3\columnwidth,0);

    \draw[blue, line width = 0.5mm] (-0.5*0.3\columnwidth,0.87*0.3\columnwidth)--(0.3\columnwidth,0);
    \draw[blue, line width = 0.5mm] (-0.5*0.3\columnwidth,0.87*0.3\columnwidth)--(-0.3\columnwidth,0);
    \draw[line width = 0.5mm] (-0.5*0.3\columnwidth,0.87*0.3\columnwidth)--(-0.5*0.3\columnwidth,-0.87*0.3\columnwidth);
    \draw[line width = 0.5mm] (-0.5*0.3\columnwidth,0.87*0.3\columnwidth)--(0.5*0.3\columnwidth,-0.87*0.3\columnwidth);

    \draw[blue, line width = 0.5mm] (-0.5*0.3\columnwidth,-0.87*0.3\columnwidth)--(0.3\columnwidth,0);
    \draw[blue, line width = 0.5mm] (-0.5*0.3\columnwidth,-0.87*0.3\columnwidth)--(-0.3\columnwidth,0);
    \draw[line width = 0.5mm] (-0.5*0.3\columnwidth,-0.87*0.3\columnwidth)--(0.5*0.3\columnwidth,-0.87*0.3\columnwidth);

    \draw[blue, line width = 0.5mm] (0.5*0.3\columnwidth,-0.87*0.3\columnwidth)--(0.3\columnwidth,0);
    \draw[blue, line width = 0.5mm] (0.5*0.3\columnwidth,-0.87*0.3\columnwidth)--(-0.3\columnwidth,0);

    \draw[red, line width = 0.5mm] (-4,-0.95)--(-3.5,-0.95);
    \node[right] at (-3.5,-0.95) {\large$\gamma$};
    \draw[blue, line width = 0.5mm] (-4,-1.5)--(-3.5,-1.5);
    \node[right] at (-3.5,-1.5) {\large$\gamma^\prime$};
    \draw[black, line width = 0.5mm] (-4,-2.05)--(-3.5,-2.05);
    \node[right] at (-3.5,-2.05) {\large$\tilde{\gamma}$};

    \draw[red, line width = 0.5mm] (0.3\columnwidth,0)--(-0.3\columnwidth,0);
        
    \fill[red, line width = 1mm] (0.3\columnwidth,0)circle(2mm);
    \fill[red, line width = 1mm] (-0.3\columnwidth,0)circle(2mm);
    \fill[blue, line width = 1mm] (0.5*0.3\columnwidth,0.87*0.3\columnwidth)circle(2mm);
    \fill[blue, line width = 1mm] (-0.5*0.3\columnwidth,0.87*0.3\columnwidth)circle(2mm);
    \fill[blue, line width = 1mm] (0.5*0.3\columnwidth,-0.87*0.3\columnwidth)circle(2mm);
    \fill[blue, line width = 1mm] (-0.5*0.3\columnwidth,-0.87*0.3\columnwidth)circle(2mm);
        
        
\end{tikzpicture}
    }
    \caption{Schematic representation of an all-to-all connected hardware of $6$ qubits. $\gamma$ is the transverse coupling between two qubits at the \textit{on} point (red edges), $\gamma^\prime$ is the transverse coupling between a qubit at the \textit{on} point and an idle qubit at the \textit{off} point (blue edges) and $\tilde{\gamma}$ is the transverse coupling between two idle qubits (black edges). Active qubits are represented in red while idle qubits are in blue.}
    \label{fig:idle}
\end{figure}

\section{Quantifying the cross-talk errors}\label{section:models}

\subsection{Average gate fidelity}

In order to estimate the cross-talk errors, we need to use a metric that quantifies the error of a perturbed quantum operation $U_p$ with respect to an ideal one $U$.
We will use, as it is most commonly done, the average gate fidelity \cite{Nielsen} following the formula
\begin{equation}\label{eq:fid}
    \mathcal{F}(U,U_p)=\frac{d F_e(U,U_p)+1}{d+1},
\end{equation}
where $d=2^N$ is the dimension of the processor and $F_e$ is the entanglement fidelity and is defined on any completely positive trace preserving map $\mathcal{E}$ as
\begin{equation}\label{eq:fe}
    F_e(U,\mathcal{E})=|\langle\psi_e|U^\dag \mathcal{E}(|\psi_e\rangle\langle\psi_e|)U|\psi_e\rangle|,
\end{equation}
where $|\psi_e\rangle$ is a maximally entangled state between the register of the operated and idle qubits and a register of ancilla qubits.
Usually $\mathcal{E}=\mathcal{G}\otimes\mathcal{I}$, where $\mathcal{I}$ is the identity over the register of the idle qubits and $\mathcal{G}$ is a non-unitary operation over the register of the active qubits.
In our study we will use instead of $\mathcal{E}$ the perturbed operation $U_p$ that acts on both the active and idle qubits.
Since we are only interested in the cross-talk errors, we disregard the effect of the decoherence and the non-unitarity of the real operation.
Finally, in order to use the formula defined in \Eqref{eq:fid}, we only need to compute the unitary $U_p=\exp(-iHt_g)$, corresponding to the short time dynamics, instead of the usually studied thermodynamic limit, of a quantum Curie-Weiss model.
In the following, we use perturbative approaches to compute the unitary $U_p$.
To simplify further the problem, we study the dynamics in a rotating frame
so that the dynamics is
 only given by the Hamiltonian $H_\mathrm{S}$ and $H^\prime$, where $H^\prime$ is redefined as
\begin{equation}
H^\prime=\sum_{i=1}^2\sum_{j=3}^{N}\left(\sigma_{-}^{(i)}\sigma_{+}^{(j)} e^{i\Delta t_g}+\sigma_{+}^{(i)}\sigma_{-}^{(j)}e^{-i\Delta t_g}\right)
\end{equation}
with $\Delta=\omega_\textit{on}-\omega_\textit{off}$.
We use the dimensionless quantity $\gamma t_g$ as time and define $m=\gamma^\prime/\gamma$ so that in the end the Hamiltonian we are using is
\begin{equation}
    H=H_\mathrm{S}+mH^\prime,
\end{equation}
where $m\ll1$ is the order parameter.
\subsection{Perturbative expansion}
Considering that $m\ll1$, we use the power series expansion of the exponential
\begin{equation}
    U_p=\sum_{k=0}^{+\infty}\frac{(-i\gamma t_g)^k\left(H_\mathrm{S}+mH^\prime\right)^k}{k!}
\end{equation}
function to approximate the perturbed unitary operator $U_p$.
Collecting all first and second order terms, we find that
\begin{equation}\label{eq:p1}
    U_p\simeq U_\mathrm{S}+mL_1(H^\prime)+m^2L_2(H_\mathrm{S},H^\prime),
\end{equation}
with
\begin{equation}
\left\{\begin{aligned}
        &L_{1}(H^{\prime})=(e^{-i\gamma t_g}-1)H^{\prime}\\
        &L_{2}(H_\mathrm{S},H^\prime)=(\alpha H_{S}+\beta)(H^{\prime})^{2}.\\
    \end{aligned}\right.
\end{equation}

For the derivation of $L_1(H^{\prime})$ and $L_2(H_\mathrm{S},H^\prime)$ and the definition of $\alpha$ and $\beta$ see \Apref{A2}.
Using \Eqref{eq:p1} in \Eqref{eq:fe}, one finds that
\begin{equation}
    F_e\simeq\left|1+\langle \psi_e|U_S^\dag\left(mL_1+m^2L_2\right)|\psi_e\rangle\right|^2.
\end{equation}
Choosing $|\psi_e\rangle=\sum_{ij...k}|ij...k\rangle|ij...k\rangle/\sqrt{2^N}$ to be a Bell state between all the $N$ qubits and a register of $N$ ancilla qubits, we find that
\begin{equation}\label{eq:F_pert_th}
    F_{e}\simeq1-x nm^{2}+(x^2+y^2)n^2\frac{m^{4}}{4},
\end{equation}
where $x$ and $y$ are defined in \Eqref{sup:x} and \Eqref{sup:y} in \Apref{A2} and only depend on the gate time $\gamma t_g$.
We recall that $m=\gamma^\prime/\gamma$ where $\gamma$ and $\gamma^\prime$, defined in \Eqref{eq:gamma}, are respectively the active-active and active-idle qubit couplings and depends explicitely on the cavity-qubit couplings and frequency detuning.
Hence, there is a clear dependence of the fidelity with the qubits' detunings and cavity couplings.
\subsection{Results and discussion}
We predicted in \Eqref{eq:F_pert_th} that at leading order the entanglement fidelity scales linearly with the number of idle qubits and quadratically with the order parameter $m$.
This scaling is also predicted through a mean field approach (see \Apref{A4} for more details) in which we average the contribution of the idle qubits over a uniform distribution on the Bloch sphere.
Each idle qubit then contributes independently to a global magnetization applied to the operated spins which is equivalent to performing a single qubit $x-\;$rotation on each active qubit, while performing the two qubit gate.
It is interesting to note that not only was it necessary to compute the second order perturbation in $U_p$ for consistency of the expansion, but also because the first order term does not contribute to any error in $F_e$.
Moreover, it seems that, generally, odd contributions in the expansion vanish as we have checked that third order contributions also vanish in the fidelity while the fourth order contributions don't. It appears to be related to the fact that odd order terms in the perturbative expansion involve an odd number of permutations of the qubits, thus always resulting in a vanishing projection of the perturbed states onto the unperturbed ones.
In general, one can show that, for a perturbation $U_p$ of a unitary $U$, as soon as
\begin{equation}
    |\langle\psi|U^\dag U_p|\psi\rangle|^2+2\mathrm{Re}\left(\langle\psi|U^\dag_SU_p|\psi\rangle\right)=0
\end{equation}
the entanglement fidelity does not show any error and $F_e=1$.
This could not happen normally for a unitary operator $U_p$ but our perturbative approach does not preserve the unitarity of $U_p$.
This is one artefact of our approach.
Another one is that the real order parameter is not $m$ but $m\sqrt{n}$ as we do not renormalize the coupling with $\sqrt{n}$ as is usually done when studying Ising models.
Hence our method only works for a reasonable number of idle qubits with respect to the parameter $m$, such that $m\sqrt{n}\ll1$.
A unitary preserving pertubative approach can be designed using the Baker-Campbell-Hausdorff formula, or more specifically the Zassenhaus formula \cite{Zass}, writing
\begin{equation}\label{eq:zass}
    U_p=U_S \ e^{-im\gamma t_gH^\prime}e^{m\frac{\gamma^2t_g^2}{2}[H_S,H^\prime]}.
\end{equation}
We checked that although this approach preserves the unitarity, it is not as precise as the method we used (see \Apref{A3} and \Figref{fig:comp}) as each exponential in the Zassenhaus expansion involves terms for each order of the series expansion.
\Figref{fig:comp} panel (b) shows a comparison between the error rate predicted from a fully simulated processor using QuTip \cite{JOHANSSON2012,JOHANSSON2013} and the one predicted from our perturbation approach and from using the Zassenhaus formula as a function of the number of idle qubits.
We see a good agreement between the initial pertubative approach and the simulations, that we use as a benchmark, up to $n=12$ (N=14) (the limit we could reach in the simulations) and that this approach always performs better than the approximation based on Zassenhaus formula, hence our focus on the former in the rest of the article.
From \Eqref{eq:F_pert_th} we can predict that the model based on the perturbative approach fails in the region where the entanglement fidelity increases with the increase of the order parameter.
Indeed, it would correspond to an increase of the average gate fidelity while the amount of parasitic operations is also increasing.
This corresponds to the condition $dF_e/d(m\sqrt{n})=0$, which is found when
\begin{equation}\label{eq:thr1}
    nm^2=\frac{2x}{x^2+y^2}.
\end{equation}
\Eqref{eq:thr1} fixes a threshold above which we can't evaluate the error rate using our model.
Numerical evaluation of \Eqref{eq:thr1} gives $nm^2\simeq 0.39$.
As an example, such a number corresponds to $n=3987$ when $m=10^{-2}$ which is a very large number compared to the number of qubits that can be connected to the same resonator in practice.
In order to evaluate the number of qubits that can be connected to the same cavity we set an error tolerance threshold $E_\mathrm{thr}$.
Such a limit will correspond at first order to an ensemble of values of $n$ and $m$ such that

\begin{equation}\label{eq:cond_opt}
    \frac{2^Nn}{2^N+1}=\frac{E_\mathrm{thr}}{xm^2}.
\end{equation}
Using the error threshold of $E_\mathrm{thr}= 10^{-3}$ which corresponds to state of the art incoherent error rate and is consistent with surface code error threshold \cite{Qerr,fowler2012,xue2022} (the most widely considered error correction scheme), we can predict a maximal value for $n$ as a function of $m$ using the Lambert $W_0$ function since $n\in\mathbb{N}$.
We find
\begin{equation}\label{eq:nsol}
    \begin{split}n&\underset{\phantom{m\ll E_\mathrm{thr}}}{=}\left\lfloor\frac{E_\mathrm{thr}}{xm^2}+\frac{W_0\left(\frac{E_\mathrm{thr}\log(2)}{4xm^2}2^{-\frac{E_\mathrm{thr}}{xm^2}}\right)}{\log(2)}\right\rfloor\\
    &\underset{\frac{m}{E_\mathrm{thr}}\rightarrow 0}{\sim}\left\lfloor\frac{E_\mathrm{thr}}{xm^2}+\frac{E_\mathrm{thr}}{4xm^2}2^{-\frac{E_\mathrm{thr}}{xm^2}}\right\rfloor
    \end{split}
\end{equation}
as depicted in \Figref{fig:pert_sol}.
Hence if we fix the value of $m$ we can deduce the number of qubits that can be connected to the same resonator.
%
%
For $m\sim10^{-2}$, which we think is reasonable in the case of a single-electrons semiconducting spin qubit, our perturbative approach predicts $N=6$, see \Figref{fig:pert_sol}, while the numerical simulation using QuTip predicts $N= 7$, see \Figref{fig:comp} panel (a).
This leads us to the conclusion that the connectivity of the processor should be limited to blocks of $7$ interconnected qubits in this situations.
\Figref{fig:pert_sol} informs us further on the sensitivity of $N$ with respect to the ratio $m$ as we see that for $m\sim4\times10^{-2}$ it is not possible anymore to have even a single idle qubit at the threshold we imposed, while for $m=10^{-3}$ hundreds of idle qubits can be contained in the cavity.
\begin{figure}
    \centering
    \includegraphics[width=\columnwidth]{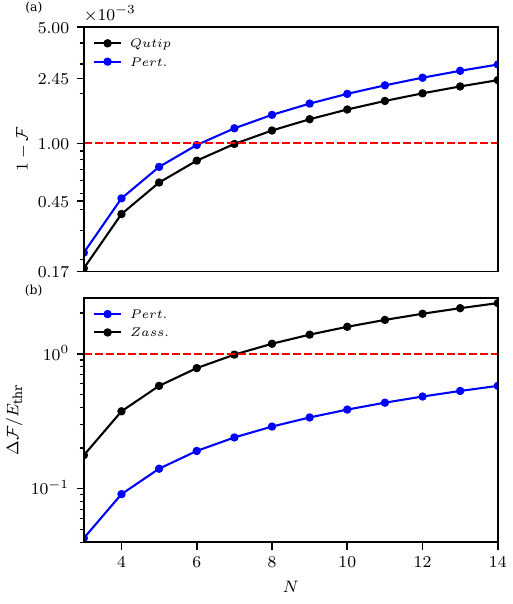}
    \caption{(a) Cross-talk error rate as a function of the number of qubits $N$ coupled to the same microwave cavity for $m=10^{-2}$ from \Eqref{eq:F_pert_th} (blue) and QuTip simulations (black). The red dashed line shows the error threshold that we set at $10^{-3}$. (b) Relative error from the approximated calculation of the error rate from \Eqref{eq:F_pert_th} (black) and \Eqref{eq:zass} (blue) with respect to the simulated error rate using QuTiP as a function of the number of qubits N coupled to the same microwave cavity for $m=10^{-2}$. The relative error is normalized by $E_\mathrm{thr}$ as we don't want a relative error bigger than the threshold we are exploring.}
\label{fig:comp}
\end{figure}
\begin{figure}[htp]
    \centering
    \includegraphics[width=\columnwidth]{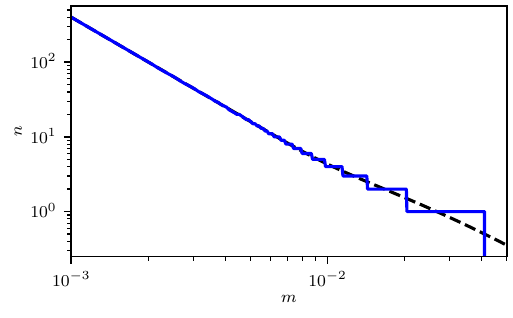}  
    \caption{Maximal number of idle qubits according to the error threshold $E_\mathrm{thr}=10^{-3}$ as a function of the ratio $m$ between the \textit{on} and \textit{off} couplings between the qubits. The plain blue line shows the solution of \Eqref{eq:nsol} while the dashed black line shows the real solutions of \Eqref{eq:cond_opt}.}
\label{fig:pert_sol}
\end{figure}
\section{Conclusion}\label{section:concl}
From the perturbative approach, we derived a scaling law for the cross-talk error rate in an all-to-all connected cQED processor tunable cavity-qubit coupling.
We predict a linear dependence of the error rate with the number of qubits in the processor and a quadratic scaling with the ratio $m$ between the \textit{on}-\textit{on} qubit transverse coupling and the \textit{on}-\textit{off} one.
From this scaling law and an error threshold that we set at $10^{-3}$, corresponding to incoherent errors, we can map for any value of $m$ a maximum number of qubits that can be connected to a single microwave resonator (see \Figref{fig:pert_sol}).
For a ratio $m\sim10^{-2}$ we find that this maximum number of qubits is $N=7$.
The quadratic scaling of the error rate with the ratio $m$ demonstrates the significant benefit of improving this ratio.
Such an improvement of the ratio $m$ would mean that an intermediate scale processor could be used with an all-to-all connectivity.
Attaining such an improvement is obviously a challenge as it requires to design a very effective \textit{off} to \textit{on} operation, requiring techniques such as optimal control \cite{Werninghaus2021}.
The connectivity limit set in our model is a strong constraint on the processor architecture and could lead to informed designs in which clusters of qubits around the same resonator are then connected together.
Obviously, other constraints such as the microwave cross-talks also limit the processor's architecture and need to be explored.
Such a reduction of the connectivity comes at the cost of an overhead in quantum algorithms corresponding to routing the qubits together in an optimal way \cite{Qubit_routing_prob} while allowing for parallel operations at the same time.
Further exploration on the hardware side needs to be done to design an optimal hardware topology.
Another possible way to reduce the effect of the cross-talks error could be to use driven two-qubit gates instead of the bare $i\mathrm{SWAP}$ as proposed previously by Heinz \textit{et al.} \cite{heinz2022} and McMillan \textit{et al.} \cite{mcmillan2022} as it could reduce the gate time and dynamically reinforce the \textit{on}-\textit{off} qubit ratio, leading to a reduction of the leakage of quantum information to the idle qubits.
Further works should focus on evaluating the advantage of using such driven gates for which the two operated qubits are directly selected from their frequencies and the driving frequency.
An alternative to the limited connectivity imposed to the processor by the cross-talk errors would be to use sub-groups of all-to-all connected qubits as logical qubits taking advantage from protected states that can be achieved in Ising $XY$ models implemented in circuit QED architecture \cite{callison2017}.
Finally, designing dynamical decoupling schemes for for addressing transverse cross-talk errors could be a way to mitigate errors in such hardware architecture and need to be explored.
This should help increase the number of qubits allowed in the all-to-all connected processor predicted in the present article. 
\begin{acknowledgments}
We thank Thomas Ayral and Grigori Matein for useful discussions leading to this project and Micheal Hynes for his review of the manuscript.
\end{acknowledgments}
\appendix
\section{Dispersive Hamiltonian}\label{A1}
In order to make appear the direct coupling between two qubits and therefore the two-qubit gate, one has to decouple the subspaces of the Hamiltonian that have different numbers of photons \cite{benito2019,warren2019}.
Using the Hamiltonian $H=H_0+H_I$, this is accomplished using a Schrieffer--Wolff transformation, consisting in a perturbative diagonalization of the Hamiltonian at second-order in the interaction term $H_I$.
In this article we define the unperturbed Hamiltonian
\begin{equation}
    H_0=\sum_{i=1}^{N}\frac{\omega_i}{2}\sigma_z^{(i)}+\omega_ca^\dag a
\end{equation}
and the interaction Hamiltonian
\begin{equation}\label{hi}
    H_I=\sum_{i=1}^{N}\left(g_i\sigma_x^{(i)}+\lambda_i\sigma_z^{(i)}\right)(a^\dag+a).
\end{equation}
In \eqref{hi}, we use a slightly more general definition of the qubit--photon interaction, introducing the longitudinal coupling $\lambda_i$, as these sort of couplings can exist in some cases.
For instance a double quantum dot (DQD) defined qubit in a circuit QED architecture shows this coupling when the DQD energy detuning is non vanishing \cite{Cottet2010,warren2019}.
We will show in the rest of this section that this does not affect the results shown in this article.
One shows that in the dispersive regime $g_i,\lambda_i\ll|\omega_c-\omega_i|$, we can define $U_{SW}=\exp(a S_+-a^\dag S_-)$, such that $S_+=S_-^\dag$ and
\begin{equation}
    S_+=\sum_{i=1}^{N}\left(\frac{\lambda_i}{\omega_c}\sigma_z^{(i)}+\eta_i\sigma_x^{(i)}+\varepsilon_i\sigma_y^{(i)}\right),
\end{equation}
where $\eta_i=g_i\omega_c/(\omega_c^2-\omega_i^2)$ and $\varepsilon_i=ig_i\omega_i/(\omega_c^2-\omega_i^2)$
We find that the transformed Hamiltonian $H_\mathrm{SW}=U^\dag_\mathrm{SW}HU_\mathrm{SW}^{\phantom\dag}$ at second order in $g_i/|\omega_c-\omega_i|$ is
\begin{equation}\label{SW}
    \begin{split}H_\mathrm{SW}=&\sum_{i=1}^{N}\left(\frac{\omega_i}{2}-g_i\eta_i\right)\sigma_z^{(i)}+\sum_{i=1,j=1,i\neq j}^{N}\left(\frac{\lambda_i\lambda_j}{\omega_j}\sigma_z^{(i)}\sigma_z^{(j)}\right.\\
    &\left.-\frac{g_i\lambda_j}{\omega_j}\sigma_z^{(j)}\sigma_x^{(i)}-g_i\eta_j\sigma_x^{(i)}\sigma_x^{(j)}-\lambda_i\eta_j\sigma_x^{(j)}\sigma_z^{(i)}\right)\end{split}
\end{equation}
where we assumed the cavity to remain in its vacuum state.
Let us first redefine the frequency of the qubit and replace $\omega_i-2g_i\eta_i$ with $\omega_i$, updating the value of $\omega_i$.
The term proportional to $\sigma_z^{(i)}\sigma_z^{(j)}$ typically appears for qubits that can be tuned electrically.
\Figref{sup:fig1} shows a typical example for single-electron semiconductor spin qubits.
We see that the terms proportional to $\sigma_z^{(i)}\sigma_z^{(j)}$ are typically orders of magnitude smaller than $m\gamma$ as $\omega_i\gg g_c$ where $g_c$ is the bare electron-photon coupling and is the maximum value that both $\lambda_i$ and $g_i$ can take.
As we neglect the terms proportional to $\tilde{\gamma}$ in the main text, for the same reason we neglect the terms proportional to $\tilde{\omega}=\lambda_i\lambda_j(\omega_i+\omega_j)/\omega_i\omega_j$.
Neglecting the fast oscillating terms in \Eqref{SW} as $\omega_i\gg \lambda_j,g_j$ for any combination of $(i,j)$, by use of a rotating wave approximation, we find the simplified Hamiltonian in \Eqref{Heff} in the main where we define $\gamma=g_i\eta_j+g_j\eta_i$ for two \textit{on} qubits, $\gamma^\prime=g_i\eta_j+g_j\eta_i$ for one \textit{off} and one \textit{on} qubit and $\tilde{\gamma}=g_i\eta_j+g_j\eta_i$ for two \textit{off} qubits. Finally we redifined the qubit frequency as $\omega_i-2g_i\eta_i$ in \Eqref{Heff} of the main text.
\begin{figure}
    \centering
    \includegraphics[width=\columnwidth]{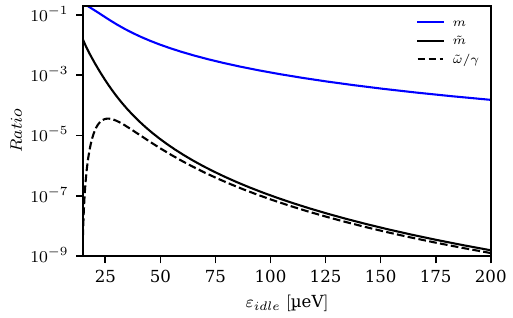}
    \caption{Example of evolution of the ratio between $\gamma^\prime$ (blue), $\tilde{\gamma}$ and $\tilde{\omega}=\lambda_i\lambda_j(\omega_i+\omega_j)/\omega_i\omega_j$ over $\gamma$ as a function of the DQD energy detuning $\varepsilon$ for single-electron semiconducting spin-qubits.}\label{sup:fig1}
\end{figure}
\section{Perturbation Theory}\label{A2}
Assuming the ability to experimentally decouple idle qubits from the microwave cavity, e.g. in the case of circuit QED based architectures by electrostatic control of the qubit, we model the cross-talk and cross-cross-talk interactions as perturbations to the iSWAP Hamiltonian
\begin{equation}\label{sup:Hfull}
\begin{split}
    H &=\gamma (H_{\mathrm{S}} + m H^{\prime} + \tilde{m}\tilde{H}) \approx \gamma (H_{\mathrm{S}} + m H^{\prime}) \\
    \end{split}
\end{equation}
where we have defined $m=\gamma^\prime/\gamma$ and $\tilde{m}=\tilde{\gamma}/\gamma$ and we have dropped the cross-cross-talk contribution (the idle-idle interactions), assuming that $m/\tilde{m}\sim 10^{-2} $ at least. 
$m$ is used as a perturbative parameter.
The perturbed propagator reads:
\begin{equation}
    \begin{split}
        U_{p} &= e^{-i\gamma t_gH_{p}}=\sum_{n=0}^{\infty} \frac{(-i\gamma t_g)^{n}}{n!} \left(H_{\mathrm{S}}+mH^\prime\right)^{n}\\
    \end{split}
\end{equation}
Since $ m \ll 1$, high order terms coming out of the binomial expansion can be neglected and the computation greatly simplifies. Arresting the expansion at second order yields to
\begin{equation}
\begin{split}    
    U_{p} &= e^{-i\gamma H_{\mathrm{S}}t_g} + m L_1\left(H_{\mathrm{S}} , H^\prime \right) + m^{2} L_2\left(H_{\mathrm{S}}, H^{\prime}\right)+ o(m^{3})\\
\end{split}
\end{equation}
where $L_1$ and $L_2$ are defined as
\begin{equation}\label{eq:first_ord_anti_comm}
    \begin{split}
   &L_1(A, B)= \sum_{n>0}^{\infty}\sum_{k=0}^{n-1} \frac{(-i\gamma t_g)^{n}A^{k}BA^{n-k-1}}{n!}\\ 
    \end{split}
\end{equation}
and
\begin{equation}\label{second_ord_anti_comm}
    \begin{split}
        &L_2(A, B)=(-i\gamma t_g)^{2}B^{2}\\
        &+\sum_{n\geq3}^{\infty}\sum_{k=0}^{n-2} \frac{(-i\gamma t_g)^{n}}{n!} A^{k} B^{2} A^{n-(k+2)}\\
        & +\sum_{n\geq3}^{\infty}\sum_{k=0}^{n-3}\sum_{l=0}^{k} \frac{(-i\gamma t_g)^{n}}{n!} A^{k-l} B A^{n-(k+2)} B A^{l}\\
    \end{split}
\end{equation}
These expressions can be simplified exploiting the properties of Pauli operators as one can show that $H_\mathrm{S}^{2n+1}=H_\mathrm{S}$, $H_\mathrm{S}^{2n}=H_\mathrm{S}^2$ and $H_\mathrm{S}(H^{\prime})^{n} H_\mathrm{S}=0$ for $n\in\mathbb{N}$.
Using these properties, we can separate \Eqref{eq:first_ord_anti_comm} and \Eqref{second_ord_anti_comm} into even and odd contribution in $n$ and we show
\begin{equation}
    L_{1}(H^\prime)=(e^{-i\gamma t_g}-1)H^{\prime}
\end{equation}
and
\begin{equation}
    L_{2}(H_\mathrm{S},H^\prime)=(\alpha H_{S}+\beta)(H^{\prime})^{2},
\end{equation}
where
\begin{equation}
\begin{aligned}
    &\alpha = c_{1}+c_{2}+c_{3}+c_{4}+2c_{5}\\
    &\beta = c_{1}+c_{2}-(\gamma t_g)^{2}\\
\end{aligned}
\end{equation}
and
\begin{equation}
    \begin{aligned}
        &c_1=\cos{(\gamma t_g)}-1+\frac{\gamma^{2}t_g^{2}}{2}-\frac{\gamma^{4}t_g^{4}}{4!}\\
        &c_2=-i\sin{(\gamma t_g)}+i\gamma t_g-\frac{i(\gamma t_g)^{3}}{3!}+\frac{i(\gamma t_g)^{5}}{5!}\\
        &c_3=2-2\cos{(\gamma t_g)}-\frac{\gamma t_g}{2}\sin{(\gamma t_g)}-\frac{(\gamma t_g)^{2}}{2}+\frac{(\gamma t_g)^{6}}{6!}\\
        &c_4=1-\cos{(\gamma t_g)}-\frac{\gamma t_g}{2}\sin{(\gamma t_g)}-\frac{(\gamma t_g)^{4}}{4!}+\frac{2(\gamma t_g)^{6}}{6!}\\
        &\begin{aligned}c_5=&-i\gamma t_g-\frac{i}{2}\gamma t_g \cos{(\gamma t_g)}+\frac{3i}{2}\sin{(\gamma t_g)}\\
        &+i\frac{(\gamma t_g)^{5}}{5!}-2i\frac{(\gamma t_g)^{7}}{7!}.\end{aligned}
    \end{aligned}
\end{equation}
In order to compute the fidelity, we only need to compute the entanglement fidelity according to \cite{Nielsen} for a maximally entangled state between the qubit register on which the gate acts and a register of ancilla qubits that are not operated by the gate.
Essentially, we define
\begin{equation}    
    \ket{\psi} = \sum_{a_{1}, a_{2},\cdots, a_{N}} \frac{1}{\sqrt{2^{N}}}\ket{a_{1}a_{2}\cdots a_{N}}\ket{a_{1} a_{2}\cdots a_{N}}
\end{equation}
where $a_{j} \in \{0,1\}$, $j \in [1,N]$ and $N$ is the number of qubits in the system.
Then the entanglement fidelity is
\begin{equation}\label{eq:Fe}
    F_e(U,\mathcal{E})=\langle\psi|U^\dag\mathcal{E}(\rho)U|\psi\rangle,
\end{equation}
where $\mathcal{E}$ is the gate we are evaluating, $U$ is the reference unitary gate $\mathcal{E}$ is compared to and $\rho$ is the density operator corresponding to state $|\psi\rangle$.
Since $\mathcal{E}=U_p$ is a unitary operator in our case, as we do not take into account any decoherence process, \Eqref{eq:Fe} simplifies into
\begin{equation}
    F_e(U_\mathrm{S},U_p)=\left|\langle\psi|U^\dag_\mathcal{S}U_p|\psi\rangle\right|^2.
\end{equation}
Using the fact that $H$ is only made of permutation operators between two qubits in the register in different states, one can easily find that the first order term $L_1(H_\mathrm{S},H^\prime)$ does not contribute to the fidelity.
Indeed, writing
\begin{equation}
    \left\{\begin{aligned}
        &H_\mathrm{S}=(1-\delta_{a_1,a_2})P(1,2)\\
        &H_\mathrm{S}^2=(1-\delta_{a_1,a_2})\mathbb{1}\\
        &H^\prime=\sum_{i=1}^{2}\sum_{j=3}^N(1-\delta_{a_i,a_j})P(i,j)e^{i(1-2\delta_{a_i,1})\Delta t},
    \end{aligned}\right.
\end{equation}
where $\mathrm{I}$ is the identity operator, $\delta_{i,j}$ is the Kronecker delta and $P(i,j)|a_1...a_i...a_j...a_N\rangle=|a_1...a_j...a_i...a_N\rangle$ is the permutation of the qubit states $|a_i\rangle$ and $|a_j\rangle$,
one shows that as $U^\dag_\mathrm{S}U_p$ will always consists in an odd number of permutations resulting in a state that is always orthogonal to the original one.
Using the same approach on the second order term $L_2(H_\mathrm{S},H^\prime)$, one find the formula in \Eqref{eq:F_pert_th} in the main text
\begin{equation}\label{sup:F_pert}
    F_e\simeq1-xnm^2+(x^2+y^2)n^2\frac{m^4}{4},
\end{equation}
where we find that
\begin{widetext}
\begin{equation}\label{sup:x}
        x  = 1-\cos{(\gamma t_g})-\sin{(\gamma t_g)}+\gamma t_g\cos{(\gamma t_g)}+\frac{(\gamma t_g)^{2}}{2}+2\frac{(\gamma t_g)^{3}}{3!}+\frac{(\gamma t_g)^{4}}{4!}-4\frac{(\gamma t_g)^{5}}{5!}+4\frac{(\gamma t_g)^{7}}{7!}
\end{equation}
and
\begin{equation}\label{sup:y}
        y  = 1-\cos{(\gamma t_g)}+\sin{(\gamma t_g)}-\gamma t_g-\gamma t_g\sin{(\gamma t_g)}-\frac{(\gamma t_g)^{2}}{2}+\frac{(\gamma t_g)^{3}}{3!}-3\frac{(\gamma t_g)^{4}}{4!}-\frac{(\gamma t_g)^{5}}{5!}+3\frac{(\gamma t_g)^{6}}{6!}.
\end{equation}
\end{widetext}
For the gate time of an $i\mathrm{SWAP}$, $\gamma t_g=\frac{\pi}{2}$, one finds $x\approx 2.47$ and $x^{2} + y^{2}\approx 12.43$.
These are the values for $x$ and $y$ that are used in the numerical application in the main text.
\section{The Zassenhaus decomposition}\label{A3}
As mentioned in the main text, while the perturbative approach gives relatively good results in comparison to the numerical simulation, one concern that we have is that it does not preserves the unitarity of the operator $U_p$.
To adress this issue, we explore an alternative expansion based on the Baker-Campbell-Hausdorff formula, more specifically on the Zassenhaus formula.
The propagator of the Hamiltonian of \Eqref{sup:Hfull} writes
\begin{equation}
    U_p = e^{-i\gamma t_g(H_{\mathrm{S}}+ mH^{\prime})}.
\end{equation}
This can be approximated via the Zassenhaus decomposition \cite{Zass}, which at first order is most commonly known as the Suzuki-Trotter decomposition \cite{Suzuki}.
Applying the Zassenhaus decomposition at second order yields
\begin{equation}
\begin{split}
&U_{Z} = e^{-i\gamma t_g H_{\mathrm{S}}}
e^{-im\gamma t_gH^{\prime}}
e^{-m\frac{(\gamma t_g)^{2}}{2}[H_{\mathrm{S}}, H^{\prime}]}
\end{split}
\end{equation}
Recalling the definition of $H^{\prime}=\sum_{i=1,j=3}^{2,N}H_{ij}$, where we defined
\begin{equation}
    H_{ij}=\sigma_-^{(i)}\sigma^{(j)}_+e^{i\Delta t_g}+\sigma_+^{(i)}\sigma_-^{(j)}e^{-\Delta t_g},
\end{equation}
we compute the cross-talk propagators
\begin{equation}
    e^{-im\gamma t_gH_{ij}}=1+[\cos{(m\gamma t_g)}-1] H_{ij}^{2} + i\sin{(m\gamma t_g)}H_{ij}.
\end{equation}
Let's now focus on  $e^{\frac{-t_g^{2}}{2}[\gamma H_{\mathrm{s}}, \gamma ^{\prime}H^{\prime}]}$.
Exploiting the properties of Pauli matrices, we have
\begin{equation}
    [H_{ij}, H_{jk}]=(-  \sigma_{-}^{(i)}\sigma_{+}^{(k)} e^{i\Delta t_g} + \sigma_{+}^{(i)} \sigma_{-}^{(k)}e^{-i\Delta t_g}) \sigma_{z}^{(j)}
\end{equation}
and the exponential of the commutator is
\begin{widetext}
\begin{equation}
        e^{-m\frac{(\gamma t_g)^{2}}{2}[H_{ij}, H_{jk}]} =1+\left[\cos{\left(m\frac{(\gamma t_g)^{2}}{4}\right)}-1 \right] H_{ik}^{2}- \sin{\left(m\frac{(\gamma t_g)^2}{4}\right)}\tilde{H}_{ik}\sigma_{z}^{(j)}.
\end{equation}
\end{widetext}
where $H_{\mathrm{S}}=H_{12}, \ k>2$ and, exploiting the properties of Pauli operators, we have defined the following, respectively for $n>0$ even and odd:
\begin{equation}
        \begin{split}
            &H_{ij}^{n}=H_{ij}^{2}=\\
            &=\frac{1}{4}\left[ (1+\sigma_{z})^{(i)} (1-\sigma_{z})^{(j)}+(1-\sigma_{z})^{(i)}(1+\sigma_{z})^{(j)}\right]\\
            \\
            &\Tilde{H}_{ij}^{n}=\Tilde{H}_{ij}:=- \sigma_{-}^{(i)} \sigma_{+}^{(j)}e^{i\Delta t} + \sigma_{+}^{(i)} \sigma_{-}^{(j)}e^{-i\Delta t}.
        \end{split}
\end{equation}
\section{Mean Field Approach}\label{A4}
In this section we present a model which has not been discussed much in the main text, but it is worth mentioning for the approach followed.
We start from the Hamiltonian of a collection of $N$ qubits
\begin{equation}
    H = \sum_{i=1}^2\frac{\omega_{i}}{2} \sigma_{z}^{(i)} + \sum_{i=1}^{N-1}\sum_{j>i}^{N}\gamma_{ij}\sigma_{x}^{(i)}\sigma_{x}^{(j)}
\end{equation}
Within the mean-Field approach, we write the idle operators as $\sigma_{k}^{(i)}= \langle \sigma_{k}^{(i)}\rangle + \delta\sigma_{k}^{(i)}$ for $k\in \{x,y,z\}$ and $i\in \{3,N\}$ , 
assuming their dynamics is
not much affected by the active qubits as their coupling is small in comparison to the inverse of the
gate time.
Neglecting the fluctuations of the idle qubits, we find the simplified Hamiltonian
\begin{equation}
    H = \sum_{i=1}^2\frac{\omega_{i}}{2} \sigma_{z}^{(i)} + H_{\mathrm{S}}+\sum_{i=1}^2\sum_{j=3}^{N}\gamma^{\prime}\langle\sigma_{x}^{(j)}\rangle\sigma_{x}^{(i)}
\end{equation}
This Hamiltonian shows that the dynamics of the gate is determined by the combination
of an $i\mathrm{SWAP}$ gate and a single qubit rotation on each active qubit where the speed of the rotations is given by $\Omega=\gamma^{\prime}\sum_{j>2}\langle\sigma_{x}^{(j)}\rangle$.
This can be understood as a total magnetization of a spin ensemble in interaction with the active qubits.
Let us first have a look at a single idle qubit. We assume the qubit is in the state
\begin{equation}
    \ket{\psi_{j}}= \frac{(\alpha_{j}\ket{0} + \beta_{j}\ket{1})}{\sqrt{\alpha_{j}^{2} + \beta_{j}^{2}}}
\end{equation}
This implies that $ \langle \sigma_{x}^{(j)} \rangle = 2 \mathrm{Re}(\alpha^{*}_{j}\beta_{j})$
and $\Omega = 2 \gamma^{\prime} \sum_{j} \mathrm{Re}(\alpha^{*}_{j}\beta_{j})$.
Considering the initial state of an idle qubit as a random variable and each idle qubit to be independent,
the total magnetization is the sum of $n$ independent variables. 
A single idle qubit is in a random
state uniformly distributed on a Bloch sphere, which means that the probability distribution of $\alpha_{j} , \beta_{j}$
is $P_{j} (\alpha_{j} , \beta_{j} ) = \frac{1}{4\pi}$. We define $m_{j} = 2\gamma^\prime \mathrm{Re}(\alpha_{j}^{*} \beta_{j})$ and counting the number of ways to obtain $m_{j}$ from a wave vector will give us its probability. 
A general state in the Bloch sphere is written as:
\begin{equation}
    \ket{\psi} = \cos{\left(\frac{\theta}{2}\right)} \ket{0} + \sin{\left(\frac{\theta}{2}\right)} e^{i\phi} \ket{1}
\end{equation}
where $\theta$ and $\phi$ are the coordinates of the wave vector on the Bloch sphere. 
This is equivalent to writing
$m_{j} = \gamma^\prime \sin{(\theta)} \cos{(\phi)}$ with $\theta \in [0, \pi]$ and $\phi \in [0, 2\pi]$.
Writing the Cartesian coordinates of the state vector
\begin{equation}
    \ket{\psi} = \begin{pmatrix}\sin{\theta} \cos{(\phi)}\\
    \sin{\theta} \sin{(\phi)}\\
    \cos{(\theta)}\end{pmatrix},
\end{equation}
we see that the magnetization is given by the $x$ component of the state vector as evidently suggested by the fact that $\sigma_x$ is the $x$ projection of the spin.
This means that the magnetization is constant over a
circle of radius $1$ in the $yz$ plane at abscissa $x$.
The length of each circle then defines the total probability of $m_{j}$.
To make this clearer, we can change the coordinates and take $\phi$ as the angle between the $yz$ plane
instead of the $xy$ plane, which just corresponds to a global rotation of the Bloch sphere. 
The wave vector is then located on the Bloch sphere as
\begin{equation}
    |\psi\rangle=\begin{pmatrix}
        \sin(\theta)\\
        \cos(\theta)\sin(\phi)\\
        \cos(\theta)\cos(\phi)
    \end{pmatrix}
\end{equation}
where now $\theta \in [-\pi/2, \pi/2]$.
This transformation amounts to changing the quantification axis to $\sigma_{x}$ instead of $\sigma_{z}$ and therefore writing a vector state in the basis $\{(\ket{0} \pm \ket{1})/\sqrt{2}\}$ instead of $\{\ket{0}, \ket{1}\}$. 
In this representation $\sin{(\theta)} = m_{j}/\gamma^{\prime}$ is the radius of the circle of magnetization $m_{j}$ and we find that
\begin{equation}
dP_{j}(m_{j}) = \frac{1}{2} \cos{(\theta)} d\theta = \frac{1}{2} dm_{j} 
\end{equation}
We recover that the density of probability of $m_{j}$ is uniform over the interval $[-\gamma^\prime, \gamma^\prime]$.
The total magnetization $\Omega$ is therefore the sum over $n$ independent uniformly distributed random variable and
the resulting probability law is an Irwin-Hall law
\begin{widetext}
\begin{equation}
    P_n(\Omega) = \frac{1}{(n-1)!} \sum_{k=0}^{n} (-1)^{k} \binom{n}{k} \left(\frac{\Omega+n\gamma^\prime}{2\gamma^\prime}-k\right)^{n-1} \mathrm{H}\left(\frac{\Omega+n\gamma^\prime}{2\gamma^\prime}-k\right)
\end{equation}
\end{widetext}
where $\mathrm{H}$ is the Heaviside function and $\Omega\in [-n\gamma^{\prime}, n\gamma^{\prime}]$.
One can diagonalize the Hamiltonian to deduce the propagator and then compute the fidelity as a function of the magnetization $F_{e}(\Omega,n)=|\bra{\psi} U_{SWAP}^{\dagger}U(\Omega,n)\ket{\psi}|^{2}$, where $\ket{\psi}$ is a maximally entangled state.
This value of the fidelity is then weighted with probability $P_n(\Omega)$. 
The average gate fidelity is then given by the integral of the magnetized Fidelity over the probability of $P_n(\Omega)$ of having a set of $n$ qubits with total magnetization $\Omega$
\begin{equation}\label{eq:F_mag}
    F(U_{\mathrm{S}}, U_p)= \int_{\mathbb{R}} F(\Omega, n) P_n(\Omega) d\Omega.
\end{equation}
\Figref{fig:delta_mag_pert} shows the comparison between the relative error in fidelity from \Eqref{eq:F_mag} and the perturbative approach in comparison to the QuTip simulation.
While it shows that the mean-field approach does not perform as good as the perturbation approach, it still predicts well the scaling law.
\begin{figure}\label{fig:delta_mag_pert}
    \centering
    \includegraphics[width=\columnwidth]{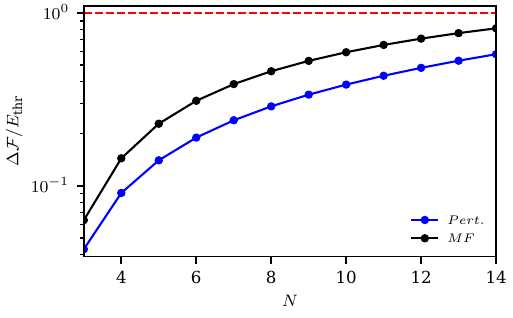}
    \caption{Relative error from the approximate calculation of the error rate from \Eqref{sup:F_pert} (black) and \Eqref{eq:F_mag} (blue) with respect to the simulated error rate using Qutip as a function of the number of qubits $N$ coupled to the same microwave cavity for $m=10^{-2}$.}
\label{fig:pert_theory_plots}
\end{figure}
%


%
\bibliographystyle{apsrev4-2}

\begin{thebibliography}{58}%
\makeatletter
\providecommand \@ifxundefined [1]{%
 \@ifx{#1\undefined}
}%
\providecommand \@ifnum [1]{%
 \ifnum #1\expandafter \@firstoftwo
 \else \expandafter \@secondoftwo
 \fi
}%
\providecommand \@ifx [1]{%
 \ifx #1\expandafter \@firstoftwo
 \else \expandafter \@secondoftwo
 \fi
}%
\providecommand \natexlab [1]{#1}%
\providecommand \enquote  [1]{``#1''}%
\providecommand \bibnamefont  [1]{#1}%
\providecommand \bibfnamefont [1]{#1}%
\providecommand \citenamefont [1]{#1}%
\providecommand \href@noop [0]{\@secondoftwo}%
\providecommand \href [0]{\begingroup \@sanitize@url \@href}%
\providecommand \@href[1]{\@@startlink{#1}\@@href}%
\providecommand \@@href[1]{\endgroup#1\@@endlink}%
\providecommand \@sanitize@url [0]{\catcode `\\12\catcode `\$12\catcode
  `\&12\catcode `\#12\catcode `\^12\catcode `\_12\catcode `\%12\relax}%
\providecommand \@@startlink[1]{}%
\providecommand \@@endlink[0]{}%
\providecommand \url  [0]{\begingroup\@sanitize@url \@url }%
\providecommand \@url [1]{\endgroup\@href {#1}{\urlprefix }}%
\providecommand \urlprefix  [0]{URL }%
\providecommand \Eprint [0]{\href }%
\providecommand \doibase [0]{https://doi.org/}%
\providecommand \selectlanguage [0]{\@gobble}%
\providecommand \bibinfo  [0]{\@secondoftwo}%
\providecommand \bibfield  [0]{\@secondoftwo}%
\providecommand \translation [1]{[#1]}%
\providecommand \BibitemOpen [0]{}%
\providecommand \bibitemStop [0]{}%
\providecommand \bibitemNoStop [0]{.\EOS\space}%
\providecommand \EOS [0]{\spacefactor3000\relax}%
\providecommand \BibitemShut  [1]{\csname bibitem#1\endcsname}%
\let\auto@bib@innerbib\@empty
\bibitem [{\citenamefont {Pita-Vidal}\ \emph {et~al.}(2024)\citenamefont
  {Pita-Vidal}, \citenamefont {Wesdorp},\ and\ \citenamefont
  {Andersen}}]{pita2024}%
  \BibitemOpen
  \bibfield  {author} {\bibinfo {author} {\bibfnamefont {M.}~\bibnamefont
  {Pita-Vidal}}, \bibinfo {author} {\bibfnamefont {J.~J.}\ \bibnamefont
  {Wesdorp}},\ and\ \bibinfo {author} {\bibfnamefont {C.~K.}\ \bibnamefont
  {Andersen}},\ }\href@noop {} {\bibfield  {journal} {\bibinfo  {journal}
  {arXiv preprint arXiv:2405.09988}\ } (\bibinfo {year} {2024})}\BibitemShut
  {NoStop}%
\bibitem [{\citenamefont {Haroche}\ and\ \citenamefont
  {Kleppner}(1989)}]{haroche1989}%
  \BibitemOpen
  \bibfield  {author} {\bibinfo {author} {\bibfnamefont {S.}~\bibnamefont
  {Haroche}}\ and\ \bibinfo {author} {\bibfnamefont {D.}~\bibnamefont
  {Kleppner}},\ }\href@noop {} {\bibfield  {journal} {\bibinfo  {journal}
  {Physics Today}\ }\textbf {\bibinfo {volume} {42}},\ \bibinfo {pages} {24}
  (\bibinfo {year} {1989})}\BibitemShut {NoStop}%
\bibitem [{\citenamefont {Tavis}\ and\ \citenamefont
  {Cummings}(1968)}]{tavis1968}%
  \BibitemOpen
  \bibfield  {author} {\bibinfo {author} {\bibfnamefont {M.}~\bibnamefont
  {Tavis}}\ and\ \bibinfo {author} {\bibfnamefont {F.~W.}\ \bibnamefont
  {Cummings}},\ }\href@noop {} {\bibfield  {journal} {\bibinfo  {journal}
  {Physical Review}\ }\textbf {\bibinfo {volume} {170}},\ \bibinfo {pages}
  {379} (\bibinfo {year} {1968})}\BibitemShut {NoStop}%
\bibitem [{\citenamefont {Chilingaryan}\ and\ \citenamefont
  {Rodr{\'\i}guez-Lara}(2013)}]{chilingaryan2013}%
  \BibitemOpen
  \bibfield  {author} {\bibinfo {author} {\bibfnamefont {S.}~\bibnamefont
  {Chilingaryan}}\ and\ \bibinfo {author} {\bibfnamefont {B.}~\bibnamefont
  {Rodr{\'\i}guez-Lara}},\ }\href@noop {} {\bibfield  {journal} {\bibinfo
  {journal} {Journal of Physics A: Mathematical and Theoretical}\ }\textbf
  {\bibinfo {volume} {46}},\ \bibinfo {pages} {335301} (\bibinfo {year}
  {2013})}\BibitemShut {NoStop}%
\bibitem [{\citenamefont {Cottet}\ \emph {et~al.}(2015)\citenamefont {Cottet},
  \citenamefont {Kontos},\ and\ \citenamefont {Dou{\c{c}}ot}}]{cottet2015}%
  \BibitemOpen
  \bibfield  {author} {\bibinfo {author} {\bibfnamefont {A.}~\bibnamefont
  {Cottet}}, \bibinfo {author} {\bibfnamefont {T.}~\bibnamefont {Kontos}},\
  and\ \bibinfo {author} {\bibfnamefont {B.}~\bibnamefont {Dou{\c{c}}ot}},\
  }\href@noop {} {\bibfield  {journal} {\bibinfo  {journal} {Physical Review
  B}\ }\textbf {\bibinfo {volume} {91}},\ \bibinfo {pages} {205417} (\bibinfo
  {year} {2015})}\BibitemShut {NoStop}%
\bibitem [{\citenamefont {Blais}\ \emph {et~al.}(2021)\citenamefont {Blais},
  \citenamefont {Grimsmo}, \citenamefont {Girvin},\ and\ \citenamefont
  {Wallraff}}]{blais2021}%
  \BibitemOpen
  \bibfield  {author} {\bibinfo {author} {\bibfnamefont {A.}~\bibnamefont
  {Blais}}, \bibinfo {author} {\bibfnamefont {A.~L.}\ \bibnamefont {Grimsmo}},
  \bibinfo {author} {\bibfnamefont {S.~M.}\ \bibnamefont {Girvin}},\ and\
  \bibinfo {author} {\bibfnamefont {A.}~\bibnamefont {Wallraff}},\ }\href@noop
  {} {\bibfield  {journal} {\bibinfo  {journal} {Reviews of Modern Physics}\
  }\textbf {\bibinfo {volume} {93}},\ \bibinfo {pages} {025005} (\bibinfo
  {year} {2021})}\BibitemShut {NoStop}%
\bibitem [{\citenamefont {Song}\ \emph {et~al.}(2017)\citenamefont {Song},
  \citenamefont {Xu}, \citenamefont {Liu}, \citenamefont {Yang}, \citenamefont
  {Zheng}, \citenamefont {Deng}, \citenamefont {Xie}, \citenamefont {Huang},
  \citenamefont {Guo}, \citenamefont {Zhang}, \citenamefont {Zhang},
  \citenamefont {Xu}, \citenamefont {Zheng}, \citenamefont {Zhu}, \citenamefont
  {Wang}, \citenamefont {Chen}, \citenamefont {Lu}, \citenamefont {Han},\ and\
  \citenamefont {Pan}}]{song2017}%
  \BibitemOpen
  \bibfield  {author} {\bibinfo {author} {\bibfnamefont {C.}~\bibnamefont
  {Song}}, \bibinfo {author} {\bibfnamefont {K.}~\bibnamefont {Xu}}, \bibinfo
  {author} {\bibfnamefont {W.}~\bibnamefont {Liu}}, \bibinfo {author}
  {\bibfnamefont {C.-p.}\ \bibnamefont {Yang}}, \bibinfo {author}
  {\bibfnamefont {S.-B.}\ \bibnamefont {Zheng}}, \bibinfo {author}
  {\bibfnamefont {H.}~\bibnamefont {Deng}}, \bibinfo {author} {\bibfnamefont
  {Q.}~\bibnamefont {Xie}}, \bibinfo {author} {\bibfnamefont {K.}~\bibnamefont
  {Huang}}, \bibinfo {author} {\bibfnamefont {Q.}~\bibnamefont {Guo}}, \bibinfo
  {author} {\bibfnamefont {L.}~\bibnamefont {Zhang}}, \bibinfo {author}
  {\bibfnamefont {P.}~\bibnamefont {Zhang}}, \bibinfo {author} {\bibfnamefont
  {D.}~\bibnamefont {Xu}}, \bibinfo {author} {\bibfnamefont {D.}~\bibnamefont
  {Zheng}}, \bibinfo {author} {\bibfnamefont {X.}~\bibnamefont {Zhu}}, \bibinfo
  {author} {\bibfnamefont {H.}~\bibnamefont {Wang}}, \bibinfo {author}
  {\bibfnamefont {Y.}~\bibnamefont {Chen}}, \bibinfo {author} {\bibfnamefont
  {C.}~\bibnamefont {Lu}}, \bibinfo {author} {\bibfnamefont {S.}~\bibnamefont
  {Han}},\ and\ \bibinfo {author} {\bibfnamefont {J.}~\bibnamefont {Pan}},\
  }\href@noop {} {\bibfield  {journal} {\bibinfo  {journal} {Physical review
  letters}\ }\textbf {\bibinfo {volume} {119}},\ \bibinfo {pages} {180511}
  (\bibinfo {year} {2017})}\BibitemShut {NoStop}%
\bibitem [{\citenamefont {Hazra}\ \emph {et~al.}(2021)\citenamefont {Hazra},
  \citenamefont {Bhattacharjee}, \citenamefont {Chand}, \citenamefont
  {Salunkhe}, \citenamefont {Gopalakrishnan}, \citenamefont {Patankar},\ and\
  \citenamefont {Vijay}}]{hazra2021}%
  \BibitemOpen
  \bibfield  {author} {\bibinfo {author} {\bibfnamefont {S.}~\bibnamefont
  {Hazra}}, \bibinfo {author} {\bibfnamefont {A.}~\bibnamefont
  {Bhattacharjee}}, \bibinfo {author} {\bibfnamefont {M.}~\bibnamefont
  {Chand}}, \bibinfo {author} {\bibfnamefont {K.~V.}\ \bibnamefont {Salunkhe}},
  \bibinfo {author} {\bibfnamefont {S.}~\bibnamefont {Gopalakrishnan}},
  \bibinfo {author} {\bibfnamefont {M.~P.}\ \bibnamefont {Patankar}},\ and\
  \bibinfo {author} {\bibfnamefont {R.}~\bibnamefont {Vijay}},\ }\href@noop {}
  {\bibfield  {journal} {\bibinfo  {journal} {Physical Review Applied}\
  }\textbf {\bibinfo {volume} {16}},\ \bibinfo {pages} {024018} (\bibinfo
  {year} {2021})}\BibitemShut {NoStop}%
\bibitem [{\citenamefont {Monroe}\ and\ \citenamefont
  {Kim}(2013)}]{monroe2013}%
  \BibitemOpen
  \bibfield  {author} {\bibinfo {author} {\bibfnamefont {C.}~\bibnamefont
  {Monroe}}\ and\ \bibinfo {author} {\bibfnamefont {J.}~\bibnamefont {Kim}},\
  }\href@noop {} {\bibfield  {journal} {\bibinfo  {journal} {Science}\ }\textbf
  {\bibinfo {volume} {339}},\ \bibinfo {pages} {1164} (\bibinfo {year}
  {2013})}\BibitemShut {NoStop}%
\bibitem [{\citenamefont {Xia}\ \emph {et~al.}(2015)\citenamefont {Xia},
  \citenamefont {Lichtman}, \citenamefont {Maller}, \citenamefont {Carr},
  \citenamefont {Piotrowicz}, \citenamefont {Isenhower},\ and\ \citenamefont
  {Saffman}}]{xia2015}%
  \BibitemOpen
  \bibfield  {author} {\bibinfo {author} {\bibfnamefont {T.}~\bibnamefont
  {Xia}}, \bibinfo {author} {\bibfnamefont {M.}~\bibnamefont {Lichtman}},
  \bibinfo {author} {\bibfnamefont {K.}~\bibnamefont {Maller}}, \bibinfo
  {author} {\bibfnamefont {A.~W.}\ \bibnamefont {Carr}}, \bibinfo {author}
  {\bibfnamefont {M.}~\bibnamefont {Piotrowicz}}, \bibinfo {author}
  {\bibfnamefont {L.}~\bibnamefont {Isenhower}},\ and\ \bibinfo {author}
  {\bibfnamefont {M.}~\bibnamefont {Saffman}},\ }\href@noop {} {\bibfield
  {journal} {\bibinfo  {journal} {Physical review letters}\ }\textbf {\bibinfo
  {volume} {114}},\ \bibinfo {pages} {100503} (\bibinfo {year}
  {2015})}\BibitemShut {NoStop}%
\bibitem [{\citenamefont {Saffman}(2016)}]{saffman2016}%
  \BibitemOpen
  \bibfield  {author} {\bibinfo {author} {\bibfnamefont {M.}~\bibnamefont
  {Saffman}},\ }\href@noop {} {\bibfield  {journal} {\bibinfo  {journal}
  {Journal of Physics B: Atomic, Molecular and Optical Physics}\ }\textbf
  {\bibinfo {volume} {49}},\ \bibinfo {pages} {202001} (\bibinfo {year}
  {2016})}\BibitemShut {NoStop}%
\bibitem [{\citenamefont {Krinner}\ \emph {et~al.}(2020)\citenamefont
  {Krinner}, \citenamefont {Lazar}, \citenamefont {Remm}, \citenamefont
  {Andersen}, \citenamefont {Lacroix}, \citenamefont {Norris}, \citenamefont
  {Hellings}, \citenamefont {Gabureac}, \citenamefont {Eichler},\ and\
  \citenamefont {Wallraff}}]{krinner2020}%
  \BibitemOpen
  \bibfield  {author} {\bibinfo {author} {\bibfnamefont {S.}~\bibnamefont
  {Krinner}}, \bibinfo {author} {\bibfnamefont {S.}~\bibnamefont {Lazar}},
  \bibinfo {author} {\bibfnamefont {A.}~\bibnamefont {Remm}}, \bibinfo {author}
  {\bibfnamefont {C.~K.}\ \bibnamefont {Andersen}}, \bibinfo {author}
  {\bibfnamefont {N.}~\bibnamefont {Lacroix}}, \bibinfo {author} {\bibfnamefont
  {G.~J.}\ \bibnamefont {Norris}}, \bibinfo {author} {\bibfnamefont
  {C.}~\bibnamefont {Hellings}}, \bibinfo {author} {\bibfnamefont
  {M.}~\bibnamefont {Gabureac}}, \bibinfo {author} {\bibfnamefont
  {C.}~\bibnamefont {Eichler}},\ and\ \bibinfo {author} {\bibfnamefont
  {A.}~\bibnamefont {Wallraff}},\ }\href@noop {} {\bibfield  {journal}
  {\bibinfo  {journal} {Physical Review Applied}\ }\textbf {\bibinfo {volume}
  {14}},\ \bibinfo {pages} {024042} (\bibinfo {year} {2020})}\BibitemShut
  {NoStop}%
\bibitem [{\citenamefont {Heinz}\ and\ \citenamefont
  {Burkard}(2021)}]{heinz2021}%
  \BibitemOpen
  \bibfield  {author} {\bibinfo {author} {\bibfnamefont {I.}~\bibnamefont
  {Heinz}}\ and\ \bibinfo {author} {\bibfnamefont {G.}~\bibnamefont
  {Burkard}},\ }\href@noop {} {\bibfield  {journal} {\bibinfo  {journal}
  {Physical Review B}\ }\textbf {\bibinfo {volume} {104}},\ \bibinfo {pages}
  {045420} (\bibinfo {year} {2021})}\BibitemShut {NoStop}%
\bibitem [{\citenamefont {Kandala}\ \emph {et~al.}(2021)\citenamefont
  {Kandala}, \citenamefont {Wei}, \citenamefont {Srinivasan}, \citenamefont
  {Magesan}, \citenamefont {Carnevale}, \citenamefont {Keefe}, \citenamefont
  {Klaus}, \citenamefont {Dial},\ and\ \citenamefont {McKay}}]{kandala2021}%
  \BibitemOpen
  \bibfield  {author} {\bibinfo {author} {\bibfnamefont {A.}~\bibnamefont
  {Kandala}}, \bibinfo {author} {\bibfnamefont {K.~X.}\ \bibnamefont {Wei}},
  \bibinfo {author} {\bibfnamefont {S.}~\bibnamefont {Srinivasan}}, \bibinfo
  {author} {\bibfnamefont {E.}~\bibnamefont {Magesan}}, \bibinfo {author}
  {\bibfnamefont {S.}~\bibnamefont {Carnevale}}, \bibinfo {author}
  {\bibfnamefont {G.}~\bibnamefont {Keefe}}, \bibinfo {author} {\bibfnamefont
  {D.}~\bibnamefont {Klaus}}, \bibinfo {author} {\bibfnamefont
  {O.}~\bibnamefont {Dial}},\ and\ \bibinfo {author} {\bibfnamefont
  {D.}~\bibnamefont {McKay}},\ }\href@noop {} {\bibfield  {journal} {\bibinfo
  {journal} {Physical Review Letters}\ }\textbf {\bibinfo {volume} {127}},\
  \bibinfo {pages} {130501} (\bibinfo {year} {2021})}\BibitemShut {NoStop}%
\bibitem [{\citenamefont {Zhao}\ \emph {et~al.}(2022)\citenamefont {Zhao},
  \citenamefont {Linghu}, \citenamefont {Li}, \citenamefont {Xu}, \citenamefont
  {Wang}, \citenamefont {Xue}, \citenamefont {Jin},\ and\ \citenamefont
  {Yu}}]{zhao2022}%
  \BibitemOpen
  \bibfield  {author} {\bibinfo {author} {\bibfnamefont {P.}~\bibnamefont
  {Zhao}}, \bibinfo {author} {\bibfnamefont {K.}~\bibnamefont {Linghu}},
  \bibinfo {author} {\bibfnamefont {Z.}~\bibnamefont {Li}}, \bibinfo {author}
  {\bibfnamefont {P.}~\bibnamefont {Xu}}, \bibinfo {author} {\bibfnamefont
  {R.}~\bibnamefont {Wang}}, \bibinfo {author} {\bibfnamefont {G.}~\bibnamefont
  {Xue}}, \bibinfo {author} {\bibfnamefont {Y.}~\bibnamefont {Jin}},\ and\
  \bibinfo {author} {\bibfnamefont {H.}~\bibnamefont {Yu}},\ }\href@noop {}
  {\bibfield  {journal} {\bibinfo  {journal} {PRX quantum}\ }\textbf {\bibinfo
  {volume} {3}},\ \bibinfo {pages} {020301} (\bibinfo {year}
  {2022})}\BibitemShut {NoStop}%
\bibitem [{\citenamefont {Wei}\ \emph {et~al.}(2022)\citenamefont {Wei},
  \citenamefont {Magesan}, \citenamefont {Lauer}, \citenamefont {Srinivasan},
  \citenamefont {Bogorin}, \citenamefont {Carnevale}, \citenamefont {Keefe},
  \citenamefont {Kim}, \citenamefont {Klaus}, \citenamefont {Landers} \emph
  {et~al.}}]{wei2022}%
  \BibitemOpen
  \bibfield  {author} {\bibinfo {author} {\bibfnamefont {K.}~\bibnamefont
  {Wei}}, \bibinfo {author} {\bibfnamefont {E.}~\bibnamefont {Magesan}},
  \bibinfo {author} {\bibfnamefont {I.}~\bibnamefont {Lauer}}, \bibinfo
  {author} {\bibfnamefont {S.}~\bibnamefont {Srinivasan}}, \bibinfo {author}
  {\bibfnamefont {D.~F.}\ \bibnamefont {Bogorin}}, \bibinfo {author}
  {\bibfnamefont {S.}~\bibnamefont {Carnevale}}, \bibinfo {author}
  {\bibfnamefont {G.}~\bibnamefont {Keefe}}, \bibinfo {author} {\bibfnamefont
  {Y.}~\bibnamefont {Kim}}, \bibinfo {author} {\bibfnamefont {D.}~\bibnamefont
  {Klaus}}, \bibinfo {author} {\bibfnamefont {W.}~\bibnamefont {Landers}},
  \emph {et~al.},\ }\href@noop {} {\bibfield  {journal} {\bibinfo  {journal}
  {Physical Review Letters}\ }\textbf {\bibinfo {volume} {129}},\ \bibinfo
  {pages} {060501} (\bibinfo {year} {2022})}\BibitemShut {NoStop}%
\bibitem [{\citenamefont {Undseth}\ \emph {et~al.}(2023)\citenamefont
  {Undseth}, \citenamefont {Xue}, \citenamefont {Mehmandoost}, \citenamefont
  {Rimbach-Russ}, \citenamefont {Eendebak}, \citenamefont {Samkharadze},
  \citenamefont {Sammak}, \citenamefont {Dobrovitski}, \citenamefont
  {Scappucci},\ and\ \citenamefont {Vandersypen}}]{undseth2023}%
  \BibitemOpen
  \bibfield  {author} {\bibinfo {author} {\bibfnamefont {B.}~\bibnamefont
  {Undseth}}, \bibinfo {author} {\bibfnamefont {X.}~\bibnamefont {Xue}},
  \bibinfo {author} {\bibfnamefont {M.}~\bibnamefont {Mehmandoost}}, \bibinfo
  {author} {\bibfnamefont {M.}~\bibnamefont {Rimbach-Russ}}, \bibinfo {author}
  {\bibfnamefont {P.~T.}\ \bibnamefont {Eendebak}}, \bibinfo {author}
  {\bibfnamefont {N.}~\bibnamefont {Samkharadze}}, \bibinfo {author}
  {\bibfnamefont {A.}~\bibnamefont {Sammak}}, \bibinfo {author} {\bibfnamefont
  {V.~V.}\ \bibnamefont {Dobrovitski}}, \bibinfo {author} {\bibfnamefont
  {G.}~\bibnamefont {Scappucci}},\ and\ \bibinfo {author} {\bibfnamefont
  {L.~M.~K.}\ \bibnamefont {Vandersypen}},\ }\href@noop {} {\bibfield
  {journal} {\bibinfo  {journal} {Physical Review Applied}\ }\textbf {\bibinfo
  {volume} {19}},\ \bibinfo {pages} {044078} (\bibinfo {year}
  {2023})}\BibitemShut {NoStop}%
\bibitem [{\citenamefont {Zhou}\ \emph {et~al.}(2023)\citenamefont {Zhou},
  \citenamefont {Sitler}, \citenamefont {Oda}, \citenamefont {Schultz},\ and\
  \citenamefont {Quiroz}}]{zhou2023}%
  \BibitemOpen
  \bibfield  {author} {\bibinfo {author} {\bibfnamefont {Z.}~\bibnamefont
  {Zhou}}, \bibinfo {author} {\bibfnamefont {R.}~\bibnamefont {Sitler}},
  \bibinfo {author} {\bibfnamefont {Y.}~\bibnamefont {Oda}}, \bibinfo {author}
  {\bibfnamefont {K.}~\bibnamefont {Schultz}},\ and\ \bibinfo {author}
  {\bibfnamefont {G.}~\bibnamefont {Quiroz}},\ }\href@noop {} {\bibfield
  {journal} {\bibinfo  {journal} {Physical Review Letters}\ }\textbf {\bibinfo
  {volume} {131}},\ \bibinfo {pages} {210802} (\bibinfo {year}
  {2023})}\BibitemShut {NoStop}%
\bibitem [{\citenamefont {Heinz}\ \emph {et~al.}(2024)\citenamefont {Heinz},
  \citenamefont {Mills}, \citenamefont {Petta},\ and\ \citenamefont
  {Burkard}}]{heinz2024}%
  \BibitemOpen
  \bibfield  {author} {\bibinfo {author} {\bibfnamefont {I.}~\bibnamefont
  {Heinz}}, \bibinfo {author} {\bibfnamefont {A.~R.}\ \bibnamefont {Mills}},
  \bibinfo {author} {\bibfnamefont {J.~R.}\ \bibnamefont {Petta}},\ and\
  \bibinfo {author} {\bibfnamefont {G.}~\bibnamefont {Burkard}},\ }\href@noop
  {} {\bibfield  {journal} {\bibinfo  {journal} {Physical Review Research}\
  }\textbf {\bibinfo {volume} {6}},\ \bibinfo {pages} {013153} (\bibinfo {year}
  {2024})}\BibitemShut {NoStop}%
\bibitem [{\citenamefont {Sarovar}\ \emph {et~al.}(2020)\citenamefont
  {Sarovar}, \citenamefont {Proctor}, \citenamefont {Rudinger}, \citenamefont
  {Young}, \citenamefont {Nielsen},\ and\ \citenamefont
  {Blume-Kohout}}]{Sarovar2020}%
  \BibitemOpen
  \bibfield  {author} {\bibinfo {author} {\bibfnamefont {M.}~\bibnamefont
  {Sarovar}}, \bibinfo {author} {\bibfnamefont {T.}~\bibnamefont {Proctor}},
  \bibinfo {author} {\bibfnamefont {K.}~\bibnamefont {Rudinger}}, \bibinfo
  {author} {\bibfnamefont {K.}~\bibnamefont {Young}}, \bibinfo {author}
  {\bibfnamefont {E.}~\bibnamefont {Nielsen}},\ and\ \bibinfo {author}
  {\bibfnamefont {R.}~\bibnamefont {Blume-Kohout}},\ }\href
  {https://doi.org/10.22331/q-2020-09-11-321} {\bibfield  {journal} {\bibinfo
  {journal} {{Quantum}}\ }\textbf {\bibinfo {volume} {4}},\ \bibinfo {pages}
  {321} (\bibinfo {year} {2020})}\BibitemShut {NoStop}%
\bibitem [{\citenamefont {Benito}\ \emph {et~al.}(2019)\citenamefont {Benito},
  \citenamefont {Petta},\ and\ \citenamefont {Burkard}}]{benito2019}%
  \BibitemOpen
  \bibfield  {author} {\bibinfo {author} {\bibfnamefont {M.}~\bibnamefont
  {Benito}}, \bibinfo {author} {\bibfnamefont {J.~R.}\ \bibnamefont {Petta}},\
  and\ \bibinfo {author} {\bibfnamefont {G.}~\bibnamefont {Burkard}},\ }\href
  {https://doi.org/10.1103/PhysRevB.100.081412} {\bibfield  {journal} {\bibinfo
   {journal} {Phys. Rev. B}\ }\textbf {\bibinfo {volume} {100}},\ \bibinfo
  {pages} {081412(R)} (\bibinfo {year} {2019})}\BibitemShut {NoStop}%
\bibitem [{\citenamefont {Matsuo}\ \emph {et~al.}(2019)\citenamefont {Matsuo},
  \citenamefont {Hattori},\ and\ \citenamefont {Yamashita}}]{matsuo2019}%
  \BibitemOpen
  \bibfield  {author} {\bibinfo {author} {\bibfnamefont {A.}~\bibnamefont
  {Matsuo}}, \bibinfo {author} {\bibfnamefont {W.}~\bibnamefont {Hattori}},\
  and\ \bibinfo {author} {\bibfnamefont {S.}~\bibnamefont {Yamashita}},\ }in\
  \href {https://doi.org/10.1109/ISCAS.2019.8702439} {\emph {\bibinfo
  {booktitle} {2019 IEEE International Symposium on Circuits and Systems
  (ISCAS)}}}\ (\bibinfo {year} {2019})\ pp.\ \bibinfo {pages}
  {1--5}\BibitemShut {NoStop}%
\bibitem [{\citenamefont {Holmes}\ \emph {et~al.}(2020)\citenamefont {Holmes},
  \citenamefont {Johri}, \citenamefont {Guerreschi}, \citenamefont {Clarke},\
  and\ \citenamefont {Matsuura}}]{Holmes_2020}%
  \BibitemOpen
  \bibfield  {author} {\bibinfo {author} {\bibfnamefont {A.}~\bibnamefont
  {Holmes}}, \bibinfo {author} {\bibfnamefont {S.}~\bibnamefont {Johri}},
  \bibinfo {author} {\bibfnamefont {G.~G.}\ \bibnamefont {Guerreschi}},
  \bibinfo {author} {\bibfnamefont {J.~S.}\ \bibnamefont {Clarke}},\ and\
  \bibinfo {author} {\bibfnamefont {A.~Y.}\ \bibnamefont {Matsuura}},\ }\href
  {https://doi.org/10.1088/2058-9565/ab73e0} {\bibfield  {journal} {\bibinfo
  {journal} {Quantum Science and Technology}\ }\textbf {\bibinfo {volume}
  {5}},\ \bibinfo {pages} {025009} (\bibinfo {year} {2020})}\BibitemShut
  {NoStop}%
\bibitem [{\citenamefont {Hashim}\ \emph {et~al.}(2022)\citenamefont {Hashim},
  \citenamefont {Rines}, \citenamefont {Omole}, \citenamefont {Naik},
  \citenamefont {Kreikebaum}, \citenamefont {Santiago}, \citenamefont {Chong},
  \citenamefont {Siddiqi},\ and\ \citenamefont {Gokhale}}]{hashim2022}%
  \BibitemOpen
  \bibfield  {author} {\bibinfo {author} {\bibfnamefont {A.}~\bibnamefont
  {Hashim}}, \bibinfo {author} {\bibfnamefont {R.}~\bibnamefont {Rines}},
  \bibinfo {author} {\bibfnamefont {V.}~\bibnamefont {Omole}}, \bibinfo
  {author} {\bibfnamefont {R.~K.}\ \bibnamefont {Naik}}, \bibinfo {author}
  {\bibfnamefont {J.~M.}\ \bibnamefont {Kreikebaum}}, \bibinfo {author}
  {\bibfnamefont {D.~I.}\ \bibnamefont {Santiago}}, \bibinfo {author}
  {\bibfnamefont {F.~T.}\ \bibnamefont {Chong}}, \bibinfo {author}
  {\bibfnamefont {I.}~\bibnamefont {Siddiqi}},\ and\ \bibinfo {author}
  {\bibfnamefont {P.}~\bibnamefont {Gokhale}},\ }\href
  {https://doi.org/10.1103/PhysRevResearch.4.033028} {\bibfield  {journal}
  {\bibinfo  {journal} {Phys. Rev. Res.}\ }\textbf {\bibinfo {volume} {4}},\
  \bibinfo {pages} {033028} (\bibinfo {year} {2022})}\BibitemShut {NoStop}%
\bibitem [{\citenamefont {Cowtan}\ \emph {et~al.}(2019)\citenamefont {Cowtan},
  \citenamefont {Dilkes}, \citenamefont {Duncan}, \citenamefont {Krajenbrink},
  \citenamefont {Simmons},\ and\ \citenamefont
  {Sivarajah}}]{Qubit_routing_prob}%
  \BibitemOpen
  \bibfield  {author} {\bibinfo {author} {\bibfnamefont {A.}~\bibnamefont
  {Cowtan}}, \bibinfo {author} {\bibfnamefont {S.}~\bibnamefont {Dilkes}},
  \bibinfo {author} {\bibfnamefont {R.}~\bibnamefont {Duncan}}, \bibinfo
  {author} {\bibfnamefont {A.}~\bibnamefont {Krajenbrink}}, \bibinfo {author}
  {\bibfnamefont {W.}~\bibnamefont {Simmons}},\ and\ \bibinfo {author}
  {\bibfnamefont {S.}~\bibnamefont {Sivarajah}},\ }in\ \href
  {https://doi.org/10.4230/LIPIcs.TQC.2019.5} {\emph {\bibinfo {booktitle}
  {14th Conference on the Theory of Quantum Computation, Communication and
  Cryptography (TQC 2019)}}},\ \bibinfo {series} {Leibniz International
  Proceedings in Informatics (LIPIcs)}, Vol.\ \bibinfo {volume} {135},\
  \bibinfo {editor} {edited by\ \bibinfo {editor} {\bibfnamefont
  {W.}~\bibnamefont {van Dam}}\ and\ \bibinfo {editor} {\bibfnamefont
  {L.}~\bibnamefont {Mancinska}}}\ (\bibinfo  {publisher} {Schloss
  Dagstuhl--Leibniz-Zentrum fuer Informatik},\ \bibinfo {address} {Dagstuhl,
  Germany},\ \bibinfo {year} {2019})\ pp.\ \bibinfo {pages}
  {5:1--5:32}\BibitemShut {NoStop}%
\bibitem [{\citenamefont {Tripathi}\ \emph {et~al.}(2022)\citenamefont
  {Tripathi}, \citenamefont {Chen}, \citenamefont {Khezri}, \citenamefont
  {Yip}, \citenamefont {Levenson-Falk},\ and\ \citenamefont
  {Lidar}}]{tripathi2022}%
  \BibitemOpen
  \bibfield  {author} {\bibinfo {author} {\bibfnamefont {V.}~\bibnamefont
  {Tripathi}}, \bibinfo {author} {\bibfnamefont {H.}~\bibnamefont {Chen}},
  \bibinfo {author} {\bibfnamefont {M.}~\bibnamefont {Khezri}}, \bibinfo
  {author} {\bibfnamefont {K.-W.}\ \bibnamefont {Yip}}, \bibinfo {author}
  {\bibfnamefont {E.}~\bibnamefont {Levenson-Falk}},\ and\ \bibinfo {author}
  {\bibfnamefont {D.~A.}\ \bibnamefont {Lidar}},\ }\href@noop {} {\bibfield
  {journal} {\bibinfo  {journal} {Physical Review Applied}\ }\textbf {\bibinfo
  {volume} {18}},\ \bibinfo {pages} {024068} (\bibinfo {year}
  {2022})}\BibitemShut {NoStop}%
\bibitem [{\citenamefont {Ezzell}\ \emph {et~al.}(2023)\citenamefont {Ezzell},
  \citenamefont {Pokharel}, \citenamefont {Tewala}, \citenamefont {Quiroz},\
  and\ \citenamefont {Lidar}}]{ezzell2023}%
  \BibitemOpen
  \bibfield  {author} {\bibinfo {author} {\bibfnamefont {N.}~\bibnamefont
  {Ezzell}}, \bibinfo {author} {\bibfnamefont {B.}~\bibnamefont {Pokharel}},
  \bibinfo {author} {\bibfnamefont {L.}~\bibnamefont {Tewala}}, \bibinfo
  {author} {\bibfnamefont {G.}~\bibnamefont {Quiroz}},\ and\ \bibinfo {author}
  {\bibfnamefont {D.~A.}\ \bibnamefont {Lidar}},\ }\href@noop {} {\bibfield
  {journal} {\bibinfo  {journal} {Physical Review Applied}\ }\textbf {\bibinfo
  {volume} {20}},\ \bibinfo {pages} {064027} (\bibinfo {year}
  {2023})}\BibitemShut {NoStop}%
\bibitem [{\citenamefont {Niu}\ \emph {et~al.}(2024)\citenamefont {Niu},
  \citenamefont {Todri-Sanial},\ and\ \citenamefont {Bronn}}]{niu2024}%
  \BibitemOpen
  \bibfield  {author} {\bibinfo {author} {\bibfnamefont {S.}~\bibnamefont
  {Niu}}, \bibinfo {author} {\bibfnamefont {A.}~\bibnamefont {Todri-Sanial}},\
  and\ \bibinfo {author} {\bibfnamefont {N.~T.}\ \bibnamefont {Bronn}},\
  }\href@noop {} {\bibfield  {journal} {\bibinfo  {journal} {Quantum Science
  and Technology}\ } (\bibinfo {year} {2024})}\BibitemShut {NoStop}%
\bibitem [{\citenamefont {Buterakos}\ \emph {et~al.}(2018)\citenamefont
  {Buterakos}, \citenamefont {Throckmorton},\ and\ \citenamefont
  {Das~Sarma}}]{buterakos2018}%
  \BibitemOpen
  \bibfield  {author} {\bibinfo {author} {\bibfnamefont {D.}~\bibnamefont
  {Buterakos}}, \bibinfo {author} {\bibfnamefont {R.~E.}\ \bibnamefont
  {Throckmorton}},\ and\ \bibinfo {author} {\bibfnamefont {S.}~\bibnamefont
  {Das~Sarma}},\ }\href@noop {} {\bibfield  {journal} {\bibinfo  {journal}
  {Physical Review B}\ }\textbf {\bibinfo {volume} {97}},\ \bibinfo {pages}
  {045431} (\bibinfo {year} {2018})}\BibitemShut {NoStop}%
\bibitem [{\citenamefont {Schuch}\ and\ \citenamefont
  {Siewert}(2003)}]{schuch2003}%
  \BibitemOpen
  \bibfield  {author} {\bibinfo {author} {\bibfnamefont {N.}~\bibnamefont
  {Schuch}}\ and\ \bibinfo {author} {\bibfnamefont {J.}~\bibnamefont
  {Siewert}},\ }\href@noop {} {\bibfield  {journal} {\bibinfo  {journal}
  {Physical Review A}\ }\textbf {\bibinfo {volume} {67}},\ \bibinfo {pages}
  {032301} (\bibinfo {year} {2003})}\BibitemShut {NoStop}%
\bibitem [{\citenamefont {Ni}\ \emph {et~al.}(2018)\citenamefont {Ni},
  \citenamefont {Rosenband},\ and\ \citenamefont {Grimes}}]{ni2018}%
  \BibitemOpen
  \bibfield  {author} {\bibinfo {author} {\bibfnamefont {K.-K.}\ \bibnamefont
  {Ni}}, \bibinfo {author} {\bibfnamefont {T.}~\bibnamefont {Rosenband}},\ and\
  \bibinfo {author} {\bibfnamefont {D.~D.}\ \bibnamefont {Grimes}},\
  }\href@noop {} {\bibfield  {journal} {\bibinfo  {journal} {Chemical science}\
  }\textbf {\bibinfo {volume} {9}},\ \bibinfo {pages} {6830} (\bibinfo {year}
  {2018})}\BibitemShut {NoStop}%
\bibitem [{\citenamefont {Warren}\ \emph
  {et~al.}(2019{\natexlab{a}})\citenamefont {Warren}, \citenamefont {Barnes},\
  and\ \citenamefont {Economou}}]{warren2019}%
  \BibitemOpen
  \bibfield  {author} {\bibinfo {author} {\bibfnamefont {A.}~\bibnamefont
  {Warren}}, \bibinfo {author} {\bibfnamefont {E.}~\bibnamefont {Barnes}},\
  and\ \bibinfo {author} {\bibfnamefont {S.~E.}\ \bibnamefont {Economou}},\
  }\bibfield  {journal} {\bibinfo  {journal} {Physical Review B}\ }\textbf
  {\bibinfo {volume} {100}},\ \href
  {https://doi.org/10.1103/PhysRevB.100.161303} {10.1103/PhysRevB.100.161303}
  (\bibinfo {year} {2019}{\natexlab{a}}),\ \Eprint
  {https://arxiv.org/abs/1902.05704} {arXiv:1902.05704} \BibitemShut {NoStop}%
\bibitem [{\citenamefont {Sung}\ \emph {et~al.}(2021)\citenamefont {Sung},
  \citenamefont {Ding}, \citenamefont {Braum{\"u}ller}, \citenamefont
  {Veps{\"a}l{\"a}inen}, \citenamefont {Kannan}, \citenamefont {Kjaergaard},
  \citenamefont {Greene}, \citenamefont {Samach}, \citenamefont {McNally},
  \citenamefont {Kim} \emph {et~al.}}]{sung2021}%
  \BibitemOpen
  \bibfield  {author} {\bibinfo {author} {\bibfnamefont {Y.}~\bibnamefont
  {Sung}}, \bibinfo {author} {\bibfnamefont {L.}~\bibnamefont {Ding}}, \bibinfo
  {author} {\bibfnamefont {J.}~\bibnamefont {Braum{\"u}ller}}, \bibinfo
  {author} {\bibfnamefont {A.}~\bibnamefont {Veps{\"a}l{\"a}inen}}, \bibinfo
  {author} {\bibfnamefont {B.}~\bibnamefont {Kannan}}, \bibinfo {author}
  {\bibfnamefont {M.}~\bibnamefont {Kjaergaard}}, \bibinfo {author}
  {\bibfnamefont {A.}~\bibnamefont {Greene}}, \bibinfo {author} {\bibfnamefont
  {G.~O.}\ \bibnamefont {Samach}}, \bibinfo {author} {\bibfnamefont
  {C.}~\bibnamefont {McNally}}, \bibinfo {author} {\bibfnamefont
  {D.}~\bibnamefont {Kim}}, \emph {et~al.},\ }\href@noop {} {\bibfield
  {journal} {\bibinfo  {journal} {Physical Review X}\ }\textbf {\bibinfo
  {volume} {11}},\ \bibinfo {pages} {021058} (\bibinfo {year}
  {2021})}\BibitemShut {NoStop}%
\bibitem [{\citenamefont {Dijkema}\ \emph {et~al.}(2023)\citenamefont
  {Dijkema}, \citenamefont {Xue}, \citenamefont {Harvey-Collard}, \citenamefont
  {Rimbach-Russ}, \citenamefont {de~Snoo}, \citenamefont {Zheng}, \citenamefont
  {Sammak}, \citenamefont {Scappucci},\ and\ \citenamefont
  {Vandersypen}}]{dijkema2023}%
  \BibitemOpen
  \bibfield  {author} {\bibinfo {author} {\bibfnamefont {J.}~\bibnamefont
  {Dijkema}}, \bibinfo {author} {\bibfnamefont {X.}~\bibnamefont {Xue}},
  \bibinfo {author} {\bibfnamefont {P.}~\bibnamefont {Harvey-Collard}},
  \bibinfo {author} {\bibfnamefont {M.}~\bibnamefont {Rimbach-Russ}}, \bibinfo
  {author} {\bibfnamefont {S.~L.}\ \bibnamefont {de~Snoo}}, \bibinfo {author}
  {\bibfnamefont {G.}~\bibnamefont {Zheng}}, \bibinfo {author} {\bibfnamefont
  {A.}~\bibnamefont {Sammak}}, \bibinfo {author} {\bibfnamefont
  {G.}~\bibnamefont {Scappucci}},\ and\ \bibinfo {author} {\bibfnamefont
  {L.~M.}\ \bibnamefont {Vandersypen}},\ }\href@noop {} {\bibfield  {journal}
  {\bibinfo  {journal} {arXiv preprint arXiv:2310.16805}\ } (\bibinfo {year}
  {2023})}\BibitemShut {NoStop}%
\bibitem [{\citenamefont {Cottet}\ and\ \citenamefont
  {Kontos}(2010)}]{Cottet2010}%
  \BibitemOpen
  \bibfield  {author} {\bibinfo {author} {\bibfnamefont {A.}~\bibnamefont
  {Cottet}}\ and\ \bibinfo {author} {\bibfnamefont {T.}~\bibnamefont
  {Kontos}},\ }\href {https://doi.org/10.1103/PhysRevLett.105.160502}
  {\bibfield  {journal} {\bibinfo  {journal} {Physical Review Letters}\
  }\textbf {\bibinfo {volume} {105}},\ \bibinfo {pages} {1} (\bibinfo {year}
  {2010})},\ \Eprint {https://arxiv.org/abs/1005.1901} {arXiv:1005.1901}
  \BibitemShut {NoStop}%
\bibitem [{\citenamefont {Gambetta}\ \emph {et~al.}(2011)\citenamefont
  {Gambetta}, \citenamefont {Houck},\ and\ \citenamefont
  {Blais}}]{gambetta2011}%
  \BibitemOpen
  \bibfield  {author} {\bibinfo {author} {\bibfnamefont {J.}~\bibnamefont
  {Gambetta}}, \bibinfo {author} {\bibfnamefont {A.~A.}\ \bibnamefont
  {Houck}},\ and\ \bibinfo {author} {\bibfnamefont {A.}~\bibnamefont {Blais}},\
  }\href@noop {} {\bibfield  {journal} {\bibinfo  {journal} {Physical review
  letters}\ }\textbf {\bibinfo {volume} {106}},\ \bibinfo {pages} {030502}
  (\bibinfo {year} {2011})}\BibitemShut {NoStop}%
\bibitem [{\citenamefont {Dicke}(1954)}]{dicke1954}%
  \BibitemOpen
  \bibfield  {author} {\bibinfo {author} {\bibfnamefont {R.~H.}\ \bibnamefont
  {Dicke}},\ }\href@noop {} {\bibfield  {journal} {\bibinfo  {journal}
  {Physical review}\ }\textbf {\bibinfo {volume} {93}},\ \bibinfo {pages} {99}
  (\bibinfo {year} {1954})}\BibitemShut {NoStop}%
\bibitem [{\citenamefont {Warren}\ \emph
  {et~al.}(2019{\natexlab{b}})\citenamefont {Warren}, \citenamefont {Barnes},\
  and\ \citenamefont {Economou}}]{Economu}%
  \BibitemOpen
  \bibfield  {author} {\bibinfo {author} {\bibfnamefont {A.}~\bibnamefont
  {Warren}}, \bibinfo {author} {\bibfnamefont {E.}~\bibnamefont {Barnes}},\
  and\ \bibinfo {author} {\bibfnamefont {S.~E.}\ \bibnamefont {Economou}},\
  }\bibfield  {journal} {\bibinfo  {journal} {Physical Review B}\ }\textbf
  {\bibinfo {volume} {100}},\ \href
  {https://doi.org/10.1103/physrevb.100.161303(R)}
  {10.1103/physrevb.100.161303(R)} (\bibinfo {year}
  {2019}{\natexlab{b}})\BibitemShut {NoStop}%
\bibitem [{\citenamefont {Burkard}\ \emph {et~al.}(2023)\citenamefont
  {Burkard}, \citenamefont {Ladd}, \citenamefont {Pan}, \citenamefont
  {Nichol},\ and\ \citenamefont {Petta}}]{burkard2021}%
  \BibitemOpen
  \bibfield  {author} {\bibinfo {author} {\bibfnamefont {G.}~\bibnamefont
  {Burkard}}, \bibinfo {author} {\bibfnamefont {T.~D.}\ \bibnamefont {Ladd}},
  \bibinfo {author} {\bibfnamefont {A.}~\bibnamefont {Pan}}, \bibinfo {author}
  {\bibfnamefont {J.~M.}\ \bibnamefont {Nichol}},\ and\ \bibinfo {author}
  {\bibfnamefont {J.~R.}\ \bibnamefont {Petta}},\ }\href
  {https://doi.org/10.1103/RevModPhys.95.025003} {\bibfield  {journal}
  {\bibinfo  {journal} {Rev. Mod. Phys.}\ }\textbf {\bibinfo {volume} {95}},\
  \bibinfo {pages} {025003} (\bibinfo {year} {2023})}\BibitemShut {NoStop}%
\bibitem [{\citenamefont {Kjaergaard}\ \emph {et~al.}(2020)\citenamefont
  {Kjaergaard}, \citenamefont {Schwartz}, \citenamefont {Braumüller},
  \citenamefont {Krantz}, \citenamefont {Wang}, \citenamefont {Gustavsson},\
  and\ \citenamefont {Oliver}}]{kjaergaard2020}%
  \BibitemOpen
  \bibfield  {author} {\bibinfo {author} {\bibfnamefont {M.}~\bibnamefont
  {Kjaergaard}}, \bibinfo {author} {\bibfnamefont {M.~E.}\ \bibnamefont
  {Schwartz}}, \bibinfo {author} {\bibfnamefont {J.}~\bibnamefont
  {Braumüller}}, \bibinfo {author} {\bibfnamefont {P.}~\bibnamefont {Krantz}},
  \bibinfo {author} {\bibfnamefont {J.~I.-J.}\ \bibnamefont {Wang}}, \bibinfo
  {author} {\bibfnamefont {S.}~\bibnamefont {Gustavsson}},\ and\ \bibinfo
  {author} {\bibfnamefont {W.~D.}\ \bibnamefont {Oliver}},\ }\href
  {https://doi.org/10.1146/annurev-conmatphys-031119-050605} {\bibfield
  {journal} {\bibinfo  {journal} {Annual Review of Condensed Matter Physics}\
  }\textbf {\bibinfo {volume} {11}},\ \bibinfo {pages} {369} (\bibinfo {year}
  {2020})},\ \Eprint
  {https://arxiv.org/abs/https://doi.org/10.1146/annurev-conmatphys-031119-050605}
  {https://doi.org/10.1146/annurev-conmatphys-031119-050605} \BibitemShut
  {NoStop}%
\bibitem [{\citenamefont {Majer}\ \emph {et~al.}(2007)\citenamefont {Majer},
  \citenamefont {Chow}, \citenamefont {Gambetta}, \citenamefont {Koch},
  \citenamefont {Johnson}, \citenamefont {Schreier}, \citenamefont {Frunzio},
  \citenamefont {Schuster}, \citenamefont {Houck}, \citenamefont {Wallraff}
  \emph {et~al.}}]{majer2007}%
  \BibitemOpen
  \bibfield  {author} {\bibinfo {author} {\bibfnamefont {J.}~\bibnamefont
  {Majer}}, \bibinfo {author} {\bibfnamefont {J.}~\bibnamefont {Chow}},
  \bibinfo {author} {\bibfnamefont {J.}~\bibnamefont {Gambetta}}, \bibinfo
  {author} {\bibfnamefont {J.}~\bibnamefont {Koch}}, \bibinfo {author}
  {\bibfnamefont {B.}~\bibnamefont {Johnson}}, \bibinfo {author} {\bibfnamefont
  {J.}~\bibnamefont {Schreier}}, \bibinfo {author} {\bibfnamefont
  {L.}~\bibnamefont {Frunzio}}, \bibinfo {author} {\bibfnamefont
  {D.}~\bibnamefont {Schuster}}, \bibinfo {author} {\bibfnamefont {A.~A.}\
  \bibnamefont {Houck}}, \bibinfo {author} {\bibfnamefont {A.}~\bibnamefont
  {Wallraff}}, \emph {et~al.},\ }\href@noop {} {\bibfield  {journal} {\bibinfo
  {journal} {Nature}\ }\textbf {\bibinfo {volume} {449}},\ \bibinfo {pages}
  {443} (\bibinfo {year} {2007})}\BibitemShut {NoStop}%
\bibitem [{\citenamefont {Bialczak}\ \emph {et~al.}(2011)\citenamefont
  {Bialczak}, \citenamefont {Ansmann}, \citenamefont {Hofheinz}, \citenamefont
  {Lenander}, \citenamefont {Lucero}, \citenamefont {Neeley}, \citenamefont
  {O’Connell}, \citenamefont {Sank}, \citenamefont {Wang}, \citenamefont
  {Weides} \emph {et~al.}}]{bialczak2011}%
  \BibitemOpen
  \bibfield  {author} {\bibinfo {author} {\bibfnamefont {R.}~\bibnamefont
  {Bialczak}}, \bibinfo {author} {\bibfnamefont {M.}~\bibnamefont {Ansmann}},
  \bibinfo {author} {\bibfnamefont {M.}~\bibnamefont {Hofheinz}}, \bibinfo
  {author} {\bibfnamefont {M.}~\bibnamefont {Lenander}}, \bibinfo {author}
  {\bibfnamefont {E.}~\bibnamefont {Lucero}}, \bibinfo {author} {\bibfnamefont
  {M.}~\bibnamefont {Neeley}}, \bibinfo {author} {\bibfnamefont
  {A.}~\bibnamefont {O’Connell}}, \bibinfo {author} {\bibfnamefont
  {D.}~\bibnamefont {Sank}}, \bibinfo {author} {\bibfnamefont {H.}~\bibnamefont
  {Wang}}, \bibinfo {author} {\bibfnamefont {M.}~\bibnamefont {Weides}}, \emph
  {et~al.},\ }\href@noop {} {\bibfield  {journal} {\bibinfo  {journal}
  {Physical review letters}\ }\textbf {\bibinfo {volume} {106}},\ \bibinfo
  {pages} {060501} (\bibinfo {year} {2011})}\BibitemShut {NoStop}%
\bibitem [{\citenamefont {Chen}\ \emph {et~al.}(2014)\citenamefont {Chen},
  \citenamefont {Neill}, \citenamefont {Roushan}, \citenamefont {Leung},
  \citenamefont {Fang}, \citenamefont {Barends}, \citenamefont {Kelly},
  \citenamefont {Campbell}, \citenamefont {Chen}, \citenamefont {Chiaro} \emph
  {et~al.}}]{chen2014}%
  \BibitemOpen
  \bibfield  {author} {\bibinfo {author} {\bibfnamefont {Y.}~\bibnamefont
  {Chen}}, \bibinfo {author} {\bibfnamefont {C.}~\bibnamefont {Neill}},
  \bibinfo {author} {\bibfnamefont {P.}~\bibnamefont {Roushan}}, \bibinfo
  {author} {\bibfnamefont {N.}~\bibnamefont {Leung}}, \bibinfo {author}
  {\bibfnamefont {M.}~\bibnamefont {Fang}}, \bibinfo {author} {\bibfnamefont
  {R.}~\bibnamefont {Barends}}, \bibinfo {author} {\bibfnamefont
  {J.}~\bibnamefont {Kelly}}, \bibinfo {author} {\bibfnamefont
  {B.}~\bibnamefont {Campbell}}, \bibinfo {author} {\bibfnamefont
  {Z.}~\bibnamefont {Chen}}, \bibinfo {author} {\bibfnamefont {B.}~\bibnamefont
  {Chiaro}}, \emph {et~al.},\ }\href@noop {} {\bibfield  {journal} {\bibinfo
  {journal} {Physical review letters}\ }\textbf {\bibinfo {volume} {113}},\
  \bibinfo {pages} {220502} (\bibinfo {year} {2014})}\BibitemShut {NoStop}%
\bibitem [{\citenamefont {Neill}\ \emph {et~al.}(2018)\citenamefont {Neill},
  \citenamefont {Roushan}, \citenamefont {Kechedzhi}, \citenamefont {Boixo},
  \citenamefont {Isakov}, \citenamefont {Smelyanskiy}, \citenamefont {Megrant},
  \citenamefont {Chiaro}, \citenamefont {Dunsworth}, \citenamefont {Arya} \emph
  {et~al.}}]{neill2018}%
  \BibitemOpen
  \bibfield  {author} {\bibinfo {author} {\bibfnamefont {C.}~\bibnamefont
  {Neill}}, \bibinfo {author} {\bibfnamefont {P.}~\bibnamefont {Roushan}},
  \bibinfo {author} {\bibfnamefont {K.}~\bibnamefont {Kechedzhi}}, \bibinfo
  {author} {\bibfnamefont {S.}~\bibnamefont {Boixo}}, \bibinfo {author}
  {\bibfnamefont {S.~V.}\ \bibnamefont {Isakov}}, \bibinfo {author}
  {\bibfnamefont {V.}~\bibnamefont {Smelyanskiy}}, \bibinfo {author}
  {\bibfnamefont {A.}~\bibnamefont {Megrant}}, \bibinfo {author} {\bibfnamefont
  {B.}~\bibnamefont {Chiaro}}, \bibinfo {author} {\bibfnamefont
  {A.}~\bibnamefont {Dunsworth}}, \bibinfo {author} {\bibfnamefont
  {K.}~\bibnamefont {Arya}}, \emph {et~al.},\ }\href@noop {} {\bibfield
  {journal} {\bibinfo  {journal} {Science}\ }\textbf {\bibinfo {volume}
  {360}},\ \bibinfo {pages} {195} (\bibinfo {year} {2018})}\BibitemShut
  {NoStop}%
\bibitem [{\citenamefont {Chayes}\ \emph {et~al.}(2008)\citenamefont {Chayes},
  \citenamefont {Crawford}, \citenamefont {Ioffe},\ and\ \citenamefont
  {Levit}}]{chayes2008}%
  \BibitemOpen
  \bibfield  {author} {\bibinfo {author} {\bibfnamefont {L.}~\bibnamefont
  {Chayes}}, \bibinfo {author} {\bibfnamefont {N.}~\bibnamefont {Crawford}},
  \bibinfo {author} {\bibfnamefont {D.}~\bibnamefont {Ioffe}},\ and\ \bibinfo
  {author} {\bibfnamefont {A.}~\bibnamefont {Levit}},\ }\href@noop {}
  {\bibfield  {journal} {\bibinfo  {journal} {Journal of Statistical Physics}\
  }\textbf {\bibinfo {volume} {133}},\ \bibinfo {pages} {131} (\bibinfo {year}
  {2008})}\BibitemShut {NoStop}%
\bibitem [{\citenamefont {Bulekov}(2021)}]{bulekov2021}%
  \BibitemOpen
  \bibfield  {author} {\bibinfo {author} {\bibfnamefont {A.}~\bibnamefont
  {Bulekov}},\ }in\ \href@noop {} {\emph {\bibinfo {booktitle} {Journal of
  Physics: Conference Series}}},\ Vol.\ \bibinfo {volume} {1740}\ (\bibinfo
  {organization} {IOP Publishing},\ \bibinfo {year} {2021})\ p.\ \bibinfo
  {pages} {012069}\BibitemShut {NoStop}%
\bibitem [{\citenamefont {Nielsen}(2002)}]{Nielsen}%
  \BibitemOpen
  \bibfield  {author} {\bibinfo {author} {\bibfnamefont {M.~A.}\ \bibnamefont
  {Nielsen}},\ }\href {https://doi.org/10.1016/s0375-9601(02)01272-0}
  {\bibfield  {journal} {\bibinfo  {journal} {Physics Letters A}\ }\textbf
  {\bibinfo {volume} {303}},\ \bibinfo {pages} {249} (\bibinfo {year}
  {2002})}\BibitemShut {NoStop}%
\bibitem [{\citenamefont {Casas}\ \emph {et~al.}(2012)\citenamefont {Casas},
  \citenamefont {Murua},\ and\ \citenamefont {Nadinic}}]{Zass}%
  \BibitemOpen
  \bibfield  {author} {\bibinfo {author} {\bibfnamefont {F.}~\bibnamefont
  {Casas}}, \bibinfo {author} {\bibfnamefont {A.}~\bibnamefont {Murua}},\ and\
  \bibinfo {author} {\bibfnamefont {M.}~\bibnamefont {Nadinic}},\ }\href
  {https://doi.org/10.1016/j.cpc.2012.06.006} {\bibfield  {journal} {\bibinfo
  {journal} {Computer Physics Communications}\ }\textbf {\bibinfo {volume}
  {183}},\ \bibinfo {pages} {2386} (\bibinfo {year} {2012})}\BibitemShut
  {NoStop}%
\bibitem [{\citenamefont {Johansson}\ \emph {et~al.}(2012)\citenamefont
  {Johansson}, \citenamefont {Nation},\ and\ \citenamefont
  {Nori}}]{JOHANSSON2012}%
  \BibitemOpen
  \bibfield  {author} {\bibinfo {author} {\bibfnamefont {J.}~\bibnamefont
  {Johansson}}, \bibinfo {author} {\bibfnamefont {P.}~\bibnamefont {Nation}},\
  and\ \bibinfo {author} {\bibfnamefont {F.}~\bibnamefont {Nori}},\ }\href@noop
  {} {\bibfield  {journal} {\bibinfo  {journal} {Computer Physics
  Communications}\ }\textbf {\bibinfo {volume} {183}},\ \bibinfo {pages} {1760}
  (\bibinfo {year} {2012})}\BibitemShut {NoStop}%
\bibitem [{\citenamefont {Johansson}\ \emph {et~al.}(2013)\citenamefont
  {Johansson}, \citenamefont {Nation},\ and\ \citenamefont
  {Nori}}]{JOHANSSON2013}%
  \BibitemOpen
  \bibfield  {author} {\bibinfo {author} {\bibfnamefont {J.}~\bibnamefont
  {Johansson}}, \bibinfo {author} {\bibfnamefont {P.}~\bibnamefont {Nation}},\
  and\ \bibinfo {author} {\bibfnamefont {F.}~\bibnamefont {Nori}},\ }\href@noop
  {} {\bibfield  {journal} {\bibinfo  {journal} {Computer Physics
  Communications}\ }\textbf {\bibinfo {volume} {184}},\ \bibinfo {pages} {1234}
  (\bibinfo {year} {2013})}\BibitemShut {NoStop}%
\bibitem [{\citenamefont {Gottesman}(2009)}]{Qerr}%
  \BibitemOpen
  \bibfield  {author} {\bibinfo {author} {\bibfnamefont {D.}~\bibnamefont
  {Gottesman}},\ }\href@noop {} {\bibinfo {title} {An introduction to quantum
  error correction and fault-tolerant quantum computation}} (\bibinfo {year}
  {2009}),\ \Eprint {https://arxiv.org/abs/0904.2557} {arXiv:0904.2557
  [quant-ph]} \BibitemShut {NoStop}%
\bibitem [{\citenamefont {Fowler}\ \emph {et~al.}(2012)\citenamefont {Fowler},
  \citenamefont {Mariantoni}, \citenamefont {Martinis},\ and\ \citenamefont
  {Cleland}}]{fowler2012}%
  \BibitemOpen
  \bibfield  {author} {\bibinfo {author} {\bibfnamefont {A.~G.}\ \bibnamefont
  {Fowler}}, \bibinfo {author} {\bibfnamefont {M.}~\bibnamefont {Mariantoni}},
  \bibinfo {author} {\bibfnamefont {J.~M.}\ \bibnamefont {Martinis}},\ and\
  \bibinfo {author} {\bibfnamefont {A.~N.}\ \bibnamefont {Cleland}},\
  }\href@noop {} {\bibfield  {journal} {\bibinfo  {journal} {Physical Review
  A}\ }\textbf {\bibinfo {volume} {86}},\ \bibinfo {pages} {032324} (\bibinfo
  {year} {2012})}\BibitemShut {NoStop}%
\bibitem [{\citenamefont {Xue}\ \emph {et~al.}(2022)\citenamefont {Xue},
  \citenamefont {Russ}, \citenamefont {Samkharadze}, \citenamefont {Undseth},
  \citenamefont {Sammak}, \citenamefont {Scappucci},\ and\ \citenamefont
  {Vandersypen}}]{xue2022}%
  \BibitemOpen
  \bibfield  {author} {\bibinfo {author} {\bibfnamefont {X.}~\bibnamefont
  {Xue}}, \bibinfo {author} {\bibfnamefont {M.}~\bibnamefont {Russ}}, \bibinfo
  {author} {\bibfnamefont {N.}~\bibnamefont {Samkharadze}}, \bibinfo {author}
  {\bibfnamefont {B.}~\bibnamefont {Undseth}}, \bibinfo {author} {\bibfnamefont
  {A.}~\bibnamefont {Sammak}}, \bibinfo {author} {\bibfnamefont
  {G.}~\bibnamefont {Scappucci}},\ and\ \bibinfo {author} {\bibfnamefont
  {L.~M.}\ \bibnamefont {Vandersypen}},\ }\href@noop {} {\bibfield  {journal}
  {\bibinfo  {journal} {Nature}\ }\textbf {\bibinfo {volume} {601}},\ \bibinfo
  {pages} {343} (\bibinfo {year} {2022})}\BibitemShut {NoStop}%
\bibitem [{\citenamefont {Werninghaus}\ \emph {et~al.}(2021)\citenamefont
  {Werninghaus}, \citenamefont {Egger}, \citenamefont {Roy}, \citenamefont
  {Machnes}, \citenamefont {Wilhelm},\ and\ \citenamefont
  {Filipp}}]{Werninghaus2021}%
  \BibitemOpen
  \bibfield  {author} {\bibinfo {author} {\bibfnamefont {M.}~\bibnamefont
  {Werninghaus}}, \bibinfo {author} {\bibfnamefont {D.~J.}\ \bibnamefont
  {Egger}}, \bibinfo {author} {\bibfnamefont {F.}~\bibnamefont {Roy}}, \bibinfo
  {author} {\bibfnamefont {S.}~\bibnamefont {Machnes}}, \bibinfo {author}
  {\bibfnamefont {F.~K.}\ \bibnamefont {Wilhelm}},\ and\ \bibinfo {author}
  {\bibfnamefont {S.}~\bibnamefont {Filipp}},\ }\href@noop {} {\bibfield
  {journal} {\bibinfo  {journal} {npj Quantum Information}\ }\textbf {\bibinfo
  {volume} {7}},\ \bibinfo {pages} {14} (\bibinfo {year} {2021})}\BibitemShut
  {NoStop}%
\bibitem [{\citenamefont {Heinz}\ and\ \citenamefont
  {Burkard}(2022)}]{heinz2022}%
  \BibitemOpen
  \bibfield  {author} {\bibinfo {author} {\bibfnamefont {I.}~\bibnamefont
  {Heinz}}\ and\ \bibinfo {author} {\bibfnamefont {G.}~\bibnamefont
  {Burkard}},\ }\href@noop {} {\bibfield  {journal} {\bibinfo  {journal}
  {Physical Review B}\ }\textbf {\bibinfo {volume} {105}},\ \bibinfo {pages}
  {L121402} (\bibinfo {year} {2022})}\BibitemShut {NoStop}%
\bibitem [{\citenamefont {McMillan}\ and\ \citenamefont
  {Burkard}(2022)}]{mcmillan2022}%
  \BibitemOpen
  \bibfield  {author} {\bibinfo {author} {\bibfnamefont {S.~R.}\ \bibnamefont
  {McMillan}}\ and\ \bibinfo {author} {\bibfnamefont {G.}~\bibnamefont
  {Burkard}},\ }\href@noop {} {\bibfield  {journal} {\bibinfo  {journal} {arXiv
  preprint arXiv:2207.13588}\ } (\bibinfo {year} {2022})}\BibitemShut {NoStop}%
\bibitem [{\citenamefont {Callison}\ \emph {et~al.}(2017)\citenamefont
  {Callison}, \citenamefont {Grosfeld},\ and\ \citenamefont
  {Ginossar}}]{callison2017}%
  \BibitemOpen
  \bibfield  {author} {\bibinfo {author} {\bibfnamefont {A.}~\bibnamefont
  {Callison}}, \bibinfo {author} {\bibfnamefont {E.}~\bibnamefont {Grosfeld}},\
  and\ \bibinfo {author} {\bibfnamefont {E.}~\bibnamefont {Ginossar}},\
  }\href@noop {} {\bibfield  {journal} {\bibinfo  {journal} {Physical Review
  B}\ }\textbf {\bibinfo {volume} {96}},\ \bibinfo {pages} {085121} (\bibinfo
  {year} {2017})}\BibitemShut {NoStop}%
\bibitem [{\citenamefont {Hatano}\ and\ \citenamefont {Suzuki}(2005)}]{Suzuki}%
  \BibitemOpen
  \bibfield  {author} {\bibinfo {author} {\bibfnamefont {N.}~\bibnamefont
  {Hatano}}\ and\ \bibinfo {author} {\bibfnamefont {M.}~\bibnamefont
  {Suzuki}},\ }in\ \href {https://doi.org/10.1007/11526216_2} {\emph {\bibinfo
  {booktitle} {Quantum Annealing and Other Optimization Methods}}}\ (\bibinfo
  {publisher} {Springer Berlin Heidelberg},\ \bibinfo {year} {2005})\ pp.\
  \bibinfo {pages} {37--68}\BibitemShut {NoStop}%
\end{thebibliography}

%

\end{document}